\documentclass[reprint,onecolumn,showpacs,nofootinbib,prd,amssymb]{revtex4}

\usepackage[usenames,dvipsnames]{xcolor}

\usepackage{amssymb, amsmath}
\usepackage{epsf}
\usepackage{bm}
\usepackage{ulem}

\usepackage{natbib}
\usepackage{graphicx}
\usepackage{cancel}

\newcommand{\aap}{Astron.\ Astrophys.}
\newcommand{\mnras}{Mon.\ Not.\ R.\ Astron.\ Soc.}

\newcommand{\apjl}{Astrophys.\ J.\ Lett.}
\newcommand{\apjs}{Astrophys.\ J.\ Suppl.\ Ser.}

\def\la{\; \raise0.3ex\hbox{$<$\kern-0.75em\raise-1.1ex\hbox{$\sim$}}\;}
\def\ga{\;  \raise0.3ex\hbox{$>$\kern-0.75em\raise-1.1ex\hbox{$\sim$}}\;}

\renewcommand{\d}[2]{\frac{d #1}{d #2}} 

\begin{document}

\title{Resonance suppression of the r-mode instability in superfluid neutron stars: Accounting for muons and entrainment}
%
\author{Elena M. Kantor, Mikhail E. Gusakov, Vasiliy A. Dommes}
\affiliation{
Ioffe Institute,
Politekhnicheskaya 26, 194021 Saint-Petersburg, Russia
}
\date{}


\begin{abstract}
We calculate the finite-temperature r-mode spectrum 
of a superfluid neutron star 
accounting for both muons in the core and the entrainment between neutrons and protons.
We show that the standard perturbation scheme, 
considering the rotation rate as an expansion parameter, 
breaks down in this case.
We develop an original perturbation scheme 
which circumvents this problem by treating both the 
perturbations
due to rotation and (weak) entrainment simultaneously.
Applying
this scheme, we 
propose 
a simple method for calculating 
the superfluid r-mode eigenfrequency in the limit of vanishing rotation rate.
We also calculate the r-mode spectrum at finite rotation rate 
for realistic microphysics input (adopting, however, the Newtonian framework 
and Cowling approximation 
when considering perturbed oscillation equations)
and show that the normal r-mode exhibits resonances 
with superfluid r-modes
at certain values of temperatures and rotation frequencies 
in the parameter range relevant to neutron stars in low-mass X-ray binaries (LMXBs).
This turns the recently suggested phenomenological model
of resonance r-mode stabilization into a quantitative theory,
capable of explaining observations.
A strong dependence of resonance rotation rates and temperatures 
on the 
neutron superfluidity model
allows us to 
constrain the latter
by confronting our calculations with the observations of neutron stars in LMXBs.
\end{abstract}

\maketitle

\section{Introduction}
\label{intro}
As it was shown in 1998 \cite{andersson98,fm98}, 
in the absence of dissipation r-modes 
(predominantly toroidal oscillations of rotating stars restored by the Coriolis force \cite{ac01}) 
are unstable with respect to radiation of gravitational waves 
at any rotation frequency of a star. 
In practice, 
r-mode instability is mostly interesting for neutron stars (NSs), 
since only for NSs do r-modes have a reasonably fast growth rate. 
Being excited, r-modes emit gravitational waves, 
which carry off angular momentum 
from the star. 
Gravitational radiation back-reaction excites the r-mode
by increasing its amplitude. 
Dissipation opposes this process. 
Calculations show that in cold NSs r-mode instability 
is effectively damped by the shear viscosity, 
while in hot NSs it is damped by the bulk viscosity. 
In slowly rotating NSs r-mode instability is 
also
effectively suppressed, 
since gravitational radiation 
back-reaction is weak at slow rotation. 
As a result, 
only warm rapidly rotating NSs 
may fall into the so called 
``instability window''
(i.e., the region of stellar temperatures and rotation rates, where r-modes are unstable).
Such NSs are observed in the low-mass X-ray binaries (LMXBs). 
Numerous observations of 
these
NSs pose a challenge, 
because modeling shows that NSs should quickly leave 
the instability window in the course of their evolution in LMXBs \cite{levin99},
since excited r-modes heat up and spin down the stars rapidly. 
Various proposals
for reconciling
theory with observations
have been discussed in the literature (see, e.g. the reviews \cite{haskell15,gg18}).
Many of them involve some exotic physics 
(e.g.\ the presence of hyperons/quarks in the NS cores), 
or make some model-dependent 
assumptions about the mechanism of nonlinear saturation of r-modes \cite{btw07,ams12,bw13,hga14,haskell15}.  

Here we shall focus on the r-mode stabilization mechanism proposed in Refs.\ \cite{gck14a,gck14b}, 
which appeals
to resonance stabilization of r-modes by superfluid (hereafter SF) modes. 
Interestingly, this mechanism 
involves 
minimal
assumptions 
about the properties of NS matter, 
such as minimal core composition (neutrons, protons and leptons) and SF of baryons. 
Neutron SF in the NS core gives rise to two independent velocity fields:
the velocity of SF neutrons and the velocity of remaining components 
(neutron Bogoliubov thermal excitations, protons, and leptons)
%
\footnote{Note that proton superconductivity does not lead to an additional independent velocity field, since protons
are coupled to other charged particles by the electromagnetic forces.}.
%
As a result, SF NSs host specific SF modes in addition to ``normal'' oscillation modes;
the latter are close analogues of oscillation modes in non-SF NSs. 
SF modes correspond to counter-motion of SF and normal fluid components, 
and hence, in contrast to normal modes, 
dissipate strongly due to powerful mutual-friction mechanism,
that tends to equalize the velocities 
of these components \cite{als84,lm00,ly03}.
In contrast to normal modes, the eigenfrequencies of SF modes 
strongly depend on the stellar temperature 
(through the temperature dependence of neutron superfluid density).
As a consequence, 
avoided-crossings of normal and SF modes take place at certain (resonance) stellar temperatures. 
Near the resonances, eigenfunctions of strongly dissipating SF modes
admix to those of normal modes and stabilize the latter. 
Modeling (within the scenario of resonance stabilization of r-modes) shows that an NS in LMXB 
should spend most of its life in the vicinity of such avoided-crossing \cite{gck14a,gck14b,kgc16,cgk17}.

Initially, this scenario was proposed as purely phenomenological one. 
To put it on a solid ground and to prove that the avoided-crossings 
of the most unstable r-mode 
with SF modes take place in the parameter range relevant to NSs in LMXBs, 
one has to calculate the r-mode spectrum for SF NSs at finite temperatures. 
This goal had been reached in the previous studies \cite{kg17,dkg19} 
under certain assumptions. 
Namely, Ref.\ \cite{kg17} calculated the r-mode spectrum for a SF neutron star 
stratified by muons, 
neglecting the entrainment between neutrons and protons 
(i.e., assuming that motion of one particle species does not induce particle current 
of another species).
Subsequent work \cite{dkg19} accounted for the entrainment effect, 
but assumed that NS core consists of neutrons, protons, and electrons only.
Here we calculate the spectrum allowing for both muons and entrainment in the core, 
and show that together they 
change
the spectrum {\it qualitatively}. 
Note that this paper 
is an extended version of the Letter \cite{kgd20}, where we 
concentrate
more on comparison of our results with the available observations and constraining the neutron
superfluidity model.

In our numerical calculations we adopt up-to-date microphysics input. 
In this sense, although we still work in the Cowling approximation 
and in the Newtonian framework 
when dealing with perturbed oscillation equations, 
the oscillation 
spectra calculated in this paper are expected to be realistic at least qualitatively, 
while the main conclusions we arrived at in this work, we believe, are robust.

In addition to
the explanation of controversial observations of NSs in LMXBs, 
the scenario of resonance r-mode stabilization proposes a new method 
to constrain the properties of superdense matter
by finding the resonance temperatures 
in the r-mode spectrum  
and confronting 
them with available observations of NSs 
in LMXBs.
Extreme sensitivity of the calculated spectra to the 
model of neutron SF
allowed us to constrain 
the latter.

The paper is organized as follows.
In Sec.\ \ref{eqB} we provide the equations describing oscillations of rotating SF NSs. 
Sec.\ \ref{sec:r-mode} discusses the expansion of these equations in the limit of slow rotation and weak entrainment. 
In Sec.\ \ref{sec:input} we describe the microphysics input that was used in our numerical calculations. 
Sec.\ \ref{sec:res} presents the results. 
Finally, in Sec.\ \ref{sec:conc} we discuss these results and conclude. Appendix contains a detailed analysis of the behavior of SF modes in the limit of vanishing rotation rate.

\section{Oscillation equations}
\label{eqB}

We consider oscillations of a slowly rotating NS
with the spin frequency $\Omega=2\pi \nu$.
Dissipation is assumed to be small and is neglected when calculating the spectra.
We adopt Cowling approximation (do not account for metric perturbations \cite{cowling41}) and work in the Newtonian framework, 
when considering perturbed hydrodynamic equations. 
In what follows we allow for muons ($\mu$) 
in the inner layers of NSs, in addition to neutrons ($n$), protons ($p$), and electrons ($e$) 
($npe\mu$-composition),
and also take into account possible
SF of baryons (neutrons and protons) in the core.
Let all the quantities depend on time $t$ as ${\rm e}^{\imath\sigma t}$ 
in the coordinate frame rotating with the star. 
Then the linearized equations governing small oscillations of SF NSs 
in that frame  
are:

(i) Euler equation
\begin{eqnarray}
-\sigma^2 {{\pmb \xi}_{b}}+2 \imath \sigma {\pmb \Omega}\times {{\pmb \xi}_{b}}=\frac{\delta w}{w^2}{\pmb \nabla} P-\frac{{\pmb \nabla} \delta P}{w}, 
\label{euler}
\end{eqnarray}
where $w=(P+\epsilon)/c^2$, $P$ is the pressure, $\epsilon$ is the energy density,
$c$ is speed of light. Note that Eq.\ (\ref{euler}) is not a purely Newtonian one, 
it respects the fact that $P$ can be comparable to $\epsilon$ in NS cores.
Here and hereafter, 
$\delta$ stands for the Euler perturbation of some thermodynamic parameter (e.g., $\delta P$). 
The Lagrangian displacement of baryons (vanishing in equilibrium),
${\pmb \xi}_{b}$, in equation (\ref{euler}) is defined as
\begin{eqnarray}
{\pmb \xi}_{b}\equiv \frac{{\pmb j}_{b}}{\imath \sigma n_{b}},
\end{eqnarray}
where $n_{b}\equiv n_{n}+n_{p}$ and ${\pmb j}_{b}\equiv {\pmb j}_{n}+{\pmb j}_{p}$ are the baryon number density and baryon current density, respectively;
$n_i$ and ${\pmb j}_i$ 
are the number density and current density of 
particle species $i=n,p,e,\mu$.

(ii) Continuity equations for baryons and leptons 
\begin{eqnarray}
\delta n_{b}+{\rm div}(n_{b} {\pmb \xi}_{b})=0 \label{cont b}, \\
\delta n_{l}+{\rm div}(n_{l} {\pmb \xi})=0 \label{cont l}.
\end{eqnarray}
Here and hereafter, 
the subscript $l=e,\mu$ refers to leptons (electrons and muons); ${\pmb \xi}\equiv {\pmb j}_{e}/(\imath \sigma n_{e})$ is the Lagrangian displacement 
of the normal liquid component 
[we assume that all the normal-matter constituents 
(i.e., leptons and baryon thermal excitations)
move with one and the same normal velocity 
due to efficient particle collisions].
If neutrons are non-SF, then ${\pmb \xi}={\pmb \xi}_{b}$,
and hydrodynamic equations become essentially the same 
as in the normal matter (even if protons are SF, see, e.g., Ref.\ \cite{ga06}). 
Bearing this in mind
we, for brevity, 
shall call ``normal'' (or ``non-SF'') the liquid 
with non-SF neutrons,
irrespective of the actual state of protons.

(iii) The ``superfluid'' equation, analogue of the Euler equation 
for SF (neutron) liquid component
\begin{eqnarray}
h \sigma^2 {\pmb z}- 2\imath h_1 \sigma {\pmb \Omega} \times {\pmb z}=c^2 n_{e} {\pmb \nabla} \delta \eta_{e}+c^2 n_{\mu} {\pmb \nabla} \delta \eta_{\mu},
\label{sfl1}
\end{eqnarray}
where ${\pmb z}\equiv{\pmb \xi}_{b}-{\pmb \xi}$ characterizes the relative Lagrangian displacement of SF and normal components;
and
$\eta_{l}\equiv \mu_{n}-\mu_{p}-\mu_{l}$ is the chemical potential imbalance (in equilibrium $\eta_{l}=0$ \cite{hpy07}).
%
Further,
\begin{eqnarray}
h=\mu_{n} n_{b}  \left[\frac{n_{b}Y_{{pp}}}{\mu_{n}(Y_{{nn}}Y_{{pp}}-Y_{{np}}^2)}-1\right], \label{beta}\\
h_1=\mu_{n} n_{b}\left(\frac{n_{b}}{Y_{nn}\mu_n+Y_{np}\mu_p}-1\right), 
\label{gamma} 
\end{eqnarray}
where $Y_{ik}=Y_{ki}$ is the relativistic symmetric entrainment matrix \cite{ga06,gkh09a,gkh09b,ghk14},
which is the analogue of the SF mass-density matrix in the non-relativistic theory \cite{ab76}.
SF equation in the form (\ref{sfl1}) 
is valid in the weak-drag regime only, 
when the interaction between the neutron vortices and normal component (e.g., electrons) is weak, 
which is a typical situation in NSs 
(see, e.g., Refs.\ \cite{mendell91,asc06}).
The above equations should be supplemented with the relation between the thermodynamic quantities,
\begin{eqnarray}
\delta n_i=\frac{\partial n_i}{\partial P} \delta P+\frac{\partial n_i}{\partial \eta_{e}} \delta \eta_{e}+\frac{\partial n_i}{\partial \eta_{\mu}} \delta \eta_{\mu}, \label{eos}
\end{eqnarray}
where again $i=n,p,e,\mu$. 
In what follows we shall use 
$P$, $\eta_{e}$ and $\eta_{\mu}$ as independent thermodynamic variables.

It is convenient to express the non-radial displacements
$\xi_{{b}\theta}$,
$\xi_{{b}\phi}$,
$z_\theta$, and $z_\phi$
as a sum of toroidal ($T_b$, $T_z$) and poloidal ($Q_b$, $Q_z$) components
\cite{saio82}:
\begin{eqnarray}
\label{eq:xib-tp}
	\xi_{{b}\theta} = \frac{\partial}{\partial \theta}Q_b(r,\theta)+\frac{\imath m T_b(r,\theta)}{{\rm sin}\theta}
,\quad
	\xi_{{b}\phi} = \frac{\imath m Q_b(r,\theta)}{{\rm sin}\theta} -\frac{\partial}{\partial \theta}T_b(r,\theta)
,\\
\label{eq:z-tp}
	z_\theta = \frac{\partial}{\partial \theta}Q_z(r,\theta)+\frac{\imath m T_z(r,\theta)}{{\rm sin}\theta}
,\quad
	z_\phi = \frac{\imath m Q_z(r,\theta)}{{\rm sin}\theta} -\frac{\partial}{\partial \theta}T_z(r,\theta),
\end{eqnarray}
where $r$ and $\theta$ are the radial distance and polar angle 
in spherical coordinate system centered at the stellar center, 
with the axis $z$ aligned with $\pmb \Omega$.
Then, following the same procedure as for non-SF stars (e.g., \cite{lf99}), 
we expand all the unknown functions
into associated Legendre polynomials with fixed $m$:
\begin{eqnarray}
\label{eq:lm-xib}
	\xi_{{b} r}(r,\theta)=\imath \sum_{l_2}\xi_{{b} r\, l_2m}(r) P_{l_2}^m(\cos\theta)
,\\
\label{eq:lm-z}
	z_{r}(r,\theta)=\imath \sum_{l_2}z_{r\, l_2m}(r) P_{l_2}^m(\cos\theta)
,\\
\label{eq:lm-Q}
	Q_b(r,\theta)=\sum_{l_2}Q_{b\,l_2m}(r) P_{l_2}^m(\cos\theta)
,\\
\label{eq:lm-Qz}
	Q_z(r,\theta)=\sum_{l_2}Q_{z\,l_2m}(r) P_{l_2}^m(\cos\theta)
,\\ 
\label{eq:lm-T}
	T_b(r,\theta)=\sum_{l_1}T_{b\,l_1m}(r) P_{l_1}^m(\cos\theta)
,\\
\label{eq:lm-Tz}
	T_z(r,\theta)=\sum_{l_1}T_{z\,l_1m}(r) P_{l_1}^m(\cos\theta)
,\\
\label{eq:lm-dP}
	\delta P(r,\theta)=\sum_{l_2}\delta P_{l_2m}(r) P_{l_2}^m(\cos\theta)
,\\
\label{eq:lm-dmu}
	\delta \eta_{{e}}(r,\theta)=\sum_{l_2}\delta \eta_{{e}\,l_2m}(r) P_{l_2}^m(\cos\theta), \\
	\label{eq:lm-dmumu}
	\delta \eta_{{\mu}}(r,\theta)=\sum_{l_2}\delta \eta_{{\mu}\,l_2m}(r) P_{l_2}^m(\cos\theta), 
\end{eqnarray}
where the summation goes over $l_1 = m+2k$ and $l_2 = m+2k+1$ ($k=0,1,2,\ldots$)
for ``odd'' modes, and over $l_1 = m+2k+1$, $l_2 = m+2k$ for ``even'' modes
\footnote{
Equations for odd and even modes completely decouple, thus odd and even modes do not mix with each other \cite{yl00}. 
Similarly, oscillation equations (and hence the oscillation modes) 
completely decouple for different values of $m$.}.

Let us consider a slowly rotating NS, 
and expand all the quantities in a power series 
in small parameter $\Omega$ 
[below we denote by $\Omega$ 
the rotation frequency
normalized to the parameter $\Omega_0\equiv \left(G M/R^3 \right)^{1/2}$, 
which is of the order of the Kepler frequency; $M$ and $R$ are the stellar mass and radius, respectively].
We are interested in 
oscillations with the eigenfrequencies $\sigma$ 
vanishing at $\Omega \rightarrow 0$. 
Thus, $\sigma$ (normalized to $\Omega_0$) in the leading order in $\Omega$
can be represented as (e.g., \cite{saio82,pbr81,lf99})
$\sigma= \sigma_0 \Omega$, 
where $\sigma_0$ does not depend on $\Omega$.

The equations, describing purely toroidal modes in the leading order in rotation, are given by:
\begin{eqnarray}
\frac{\partial}{\partial \theta}\left({\rm sin}\theta \xi_{{b}\theta}^0\right)+\imath m \xi_{{b}\phi}^0=0, \label{contb0} \\
\sigma_0 \xi_{{b}\theta}^0+2\imath{\rm cos}\theta \xi_{{b}\phi}^0=-\frac{\imath}{m}\frac{\partial}{\partial \theta}\left[{\rm sin}\theta \left(\sigma_0 \xi_{{b}\phi}^0-2\imath {\rm cos}\theta \xi_{{b}\theta}^0\right)\right], \label{Euler0} \\
\frac{\partial}{\partial \theta}\left({\rm sin}\theta z_\theta^0\right)+\imath m z_\phi^0=0, \label{contl0}\\
h(r)\sigma_0 z_\theta^0+2\imath h_1(r){\rm cos}\theta z_\phi^0=-\frac{\imath}{m}\frac{\partial}{\partial \theta}\left\{{\rm sin}\theta \left[h(r)\sigma_0 z_\phi^0-2\imath h_1(r){\rm cos}\theta z_\theta^0\right]\right\}.  \label{sfl0general}
\end{eqnarray}
Here the index $0$ indicates the leading-order term in the expansion of eigenfunctions in the power series 
in $\Omega$. 
The first couple of equations, 
Eqs.\ (\ref{contb0}) and (\ref{Euler0}), describe the normal r-modes, 
analogous to ordinary r-modes in non-SF NSs, 
while Eqs.\ (\ref{contl0}) and (\ref{sfl0general}) 
describe SF modes driven by the relative motion 
(represented by the vector ${\pmb z}$)
of SF and normal (non-SF) liquid components \cite{ac01,ly03,agh09,kg17}. 
The solution to these two systems of equations 
allow us to determine 
the eigenfrequencies 
\begin{eqnarray}
\sigma_0=\frac{2m}{l(l+1)},
\label{sigmanorm}\\
\sigma_0=\frac{2m}{l(l+1)}\frac{h_1(r)}{h(r)}
\label{sigma0general}
\end{eqnarray}
and eigenfunctions
\begin{eqnarray}
\xi_{{b}\theta}^0=\imath m T^0_{b\,lm}(r)\frac{P_l^m({\rm cos}\theta)}{{\rm sin}\theta},\;\;\;\;\xi_{{b}\phi}^0=- T^0_{b\,lm}(r)\frac{d}{d\theta}P_l^m({\rm cos}\theta), \label{xib0}\\
z_\theta^0=\imath m T^0_{z\,lm}(r)\frac{P_l^m({\rm cos}\theta)}{{\rm sin}\theta},\;\;\;\;z_\phi^0=- T^0_{z\,lm}(r)\frac{d}{d\theta}P_l^m({\rm cos}\theta) \label{z0}
\end{eqnarray}
of normal and SF modes, respectively.

Since the function $h_1(r)/h(r)$ in Eq.\ (\ref{sigma0general}),
generally, varies throughout the star, 
the frequency (\ref{sigma0general}) 
cannot be a global oscillation frequency 
-- each stellar layer has its own different eigenfrequency
%
\footnote{
Except for some special cases when $h_1(r)/h(r)$ is constant throughout the core.}. 
%
However, if we assume that $Y_{np}=0$, then $h_1(r)=h(r)$ 
[see equations (\ref{beta})--(\ref{gamma})] 
and the eigenfrequency (\ref{sigma0general}) of SF modes reduces to 
\begin{eqnarray}
\sigma_0=\frac{2m}{l(l+1)}, 
\label{sigmasfl}
\end{eqnarray}
becoming a global solution, independent of $r$ \cite{ac01,ly03,agh09,kg17}.

As discussed in Ref.\ \cite{kg17}, for vanishing entrainment ($Y_{np}=0$) in the lowest order in rotation, 
purely toroidal modes are only possible with $l=m$. 
For a given $m$ the authors of Ref.\ \cite{kg17} found
one normal nodeless $r$-mode and an infinite set of SF $r$-modes,
all having the same $\sigma_0 = 2/(m+1)$
\cite{ac01,ly03,agh09,kg17}. 

However, when neutron and proton SFs co-exist somewhere in an NS, 
entrainment should be accounted for. 
Below we shall allow for entrainment ($Y_{np}\neq0$) by considering it as a small perturbing parameter.

\section{r-modes in the limit of weak entrainment}
\label{sec:r-mode}

\subsection{Failed attempt}
\label{subs:failed}

Assuming that the entrainment is weak, let us try to develop a
perturbation scheme in small parameter $\Delta h \equiv h/h_1 - 1$ 
($\Delta h\rightarrow 0$ at $Y_{np}\rightarrow 0$) 
in the leading order in rotation frequency.
Our aim is to
find the first-order corrections in $\Delta h$ to the eigenfrequency 
of $r$-modes 
in a SF $npe\mu$ NS. 
This approach
is analogous to that of Ref.\ \cite{dkg19}, 
where it was shown
that in a SF NS with $npe$-core $r$-modes can be calculated analytically
in the first order in $\Delta h$
%
\footnote{Note, however, that Ref.\ \cite{dkg19} defined $\Delta h$ as $\Delta h \equiv h_1/h - 1$.}.
%

Below all the quantities are taken in the leading order in rotation. 
We denote the zeroth-order in $\Delta h$ with the index $0$,
and the first-order in $\Delta h$ -- with the index $1$.
Using this notation, the eigenfrequency and eigenfunctions can be expanded in Taylor series in $\Delta h$ as:
\begin{eqnarray}
	\sigma 
	= (\sigma_{0}  + \sigma_{1})\Omega 
	= \left(\frac{2}{m+1} + \sigma_{1}\right)\Omega  \label{exp_start}
,\\
	\xi_{{b} r}= \xi_{{b} r}^{1}  
,\quad
	T_b = T^{0}_b + T^{1}_b   
,\quad
	Q_b = Q^{1}_b  
,\quad
	z_r = z_r^{1}  
,\quad
	T_z = T_z^{0} + T_z^{1}  
,\quad
	Q_z = Q_z^{1} 
,\\
	\delta P =\delta P^{0}  \Omega^2
,\quad
	\delta \eta_{e} = \delta \eta_{e}^{0}  \Omega^2
	,\quad
	\delta \eta_{\mu} = \delta \eta_{\mu}^{0} \Omega^2 \label{exp_end}
.
\end{eqnarray}

Since in the absence of entrainment the SF and normal $r$-modes are purely toroidal,
the radial and poloidal displacements in the zeroth order vanish,
$\xi_{{b} r}^{0} = z_r^{0} = Q^{0}_b = Q_z^{0} = 0$. 
For rotational modes with $\sigma \propto \Omega$ at $\Omega\rightarrow 0$, the Euler perturbation 
of any (scalar) thermodynamic parameter $f$ 
(e.g., $P$, $\mu_{l}$, etc.)
is proportional to $\Omega^2$ in the leading order in rotation and in the absence of entrainment (e.g., \cite{pbr81,lf99,lm00}). Following Ref.\ \cite{dkg19}, we assume that non-vanishing 
entrainment does not change this ordering. We will need below only leading-order terms in the expansions of scalar quantities
in the rotation frequency $\Omega$.

In the zeroth order in $\Delta h$ (i.e., at vanishing entrainment),
as discussed above, 
the eigenfrequency $\sigma_{0}$ of any r-mode equals
\begin{eqnarray}
	\sigma_{0} = \frac{2}{m+1} \label{sigma0}
,
\end{eqnarray}
and the toroidal displacements are proportional to the $l=m$ associated Legendre polynomial \cite{ac01,ly03,agh09,kg17},
\begin{eqnarray}
	T^{0}_b = T_{b\;mm}^{0} (r) P_m^m (\cos\theta) \label{T}
,\quad
	T_z^{0} = T_{z\;mm}^{0} (r) P_m^m (\cos\theta) 
.
\end{eqnarray}
The functions $T_{b\;mm}^{0} (r)$ and $T_{z\;mm}^{0} (r)$ cannot be found explicitly in the leading order in entrainment and rotation frequency. 
Assuming vanishing entrainment, Ref.\ \cite{kg17} proceeded to the next-to-leading order in rotation to calculate $T_{b\;mm}^{0} (r)$ and $T_{z\;mm}^{0} (r)$. 
In contrast,
in Ref.\ \cite{dkg19} we
worked
in the leading order in rotation, 
and accounted 
for the next-to-leading order 
terms
in the entrainment to determine these functions. 
Here we follow the approach of Ref.\ \cite{dkg19}.
To find the eigenfrequency correction $\sigma_{1}$ and the functions $T_{b\;mm}^{0} (r)$ and $T_{z\;mm}^{0} (r)$,
we shall consider the continuity equations \eqref{cont b}--\eqref{cont l},
as well as $r$, $\phi$, and $\theta$-components of the Euler equation \eqref{euler} and the SF equation \eqref{sfl1}.
The $\theta$-component of the Euler equation (combined with its $\phi$-component) reads, in the first order
in $\Delta h$
(ignoring quadratically small terms
such as $\sigma_{1} \xi_{{b}\theta}^{1}$),
\begin{eqnarray}
\label{eq:euler-th-11}
	 \sigma_{1} \xi_{{b}\theta}^{0}
	+ \sigma_{0} \xi_{{b}\theta}^{1}
	+ 2 \imath \cos\theta \xi_{{b}\phi}^{1}
	= \frac{1}{\imath m} \frac{\partial}{\partial\theta} 
	\left\{\sin\theta
	\left[
		 \sigma_{1} \xi_{{b}\phi}^{0}
		+ \sigma_{0} \xi_{{b}\phi}^{1}
		- 2 \imath \left( \xi_{{b}r}^{1} \sin\theta + \xi_{{b}\theta}^{1} \cos\theta \right) 
	\right]\right\}
.
\end{eqnarray}
Substituting relations \eqref{eq:xib-tp}, \eqref{eq:lm-xib}, \eqref{eq:lm-Q}, \eqref{eq:lm-T}
into equation~\eqref{eq:euler-th-11} divided by $\sin\theta$
and equating coefficients at the terms proportional to $P_m^m$,
one can express the function $Q_{b\;m+1,m}^{1} (r)$ through $\xi_{{b} r\, m+1,m}^{1} (r)$ and $T_{b\;mm}^{0} (r)$.
Similarly, using
the $\theta$- and $\phi$-components of the SF equation,
\begin{eqnarray}
\label{eq:sfl-th-11}
\sigma_{1} z_\theta^{0}+\sigma_{0} z_\theta^{1}+\Delta h \sigma_{0} z_\theta^{0}+2 \imath \cos\theta z_\phi^{1}
=\frac{1}{\imath m} \frac{\partial}{\partial\theta} 
	\left\{\sin\theta
	\left[
		\sigma_{1} z_\phi^{0}	+ \sigma_{0} z_\phi^{1}	+ \Delta h \sigma_{0} z_\phi^{0}	- 2 \imath \left( z_r^{1} \sin\theta + z_\theta^{1} \cos\theta \right)
	\right]\right\}
,
\end{eqnarray}
one can obtain an algebraic relation between
$Q_{z \,m+1,m}^{1} (r)$, $z_{r\, m+1,m}^{1} (r)$, and $T_{z \,mm}^{0} (r)$.

Now, taking the coefficient at $P_{m+1}^m$ in the continuity equation for baryons,
\begin{eqnarray}
	\frac{1}{n_{b}} \frac{1}{r^2}\frac{\partial}{\partial r}\left(r^2 n_{b} \xi_{{b} r}^{1}\right)
	+ \frac{1}{r\sin\theta}
	\left[
		\frac{\partial}{\partial\theta}\left(\sin\theta \frac{\partial Q^{1}_b}{\partial \theta}\right)
		- \frac{m^2 Q^{1}_b}{\sin\theta}
	\right]
	= 0
,
\end{eqnarray}
and
expressing $Q^{1}_{b\;m+1,m}$ through $T^{0}_{b\;mm}$ and $\xi_{{b}r, {m+1,m}}^{1}$,
we get the first-order ODE for $\xi_{{b}r, {m+1,m}}^{1}$:
\begin{eqnarray}
	\d{}{r} \xi_{{b} r, m+1,m}^{1}
	+\left(\frac{n_{b}'}{n_{b}}+\frac{3+m}{r}\right) \xi_{{b} r, m+1,m}^{1}
	- \frac{(1 + m)^2 (3 + 2 m) \sigma_{1}}{(2+4m)r}T^{0}_{b\;mm}
	= 0   \label{xibeq}
,
\end{eqnarray}
where the prime denotes the derivative with respect to $r$.

The continuity equations for electrons and muons, 
\begin{eqnarray}
\frac{1}{n_{e}} \frac{1}{r^2}\frac{\partial}{\partial r}\left[r^2 n_{e} \left( \xi_{{b}r}^{1} - z_r^{1} \right)\right]
		+ \frac{1}{r\sin\theta}
	\left\{
		\frac{\partial}{\partial\theta} \left[\sin\theta \left( \frac{\partial Q^{1}_b}{\partial \theta} -\frac{\partial Q_z^{1}}{\partial \theta}\right)\right]
		- \frac{m^2 (Q^{1}_b-Q_z^{1})}{\sin\theta}
	\right\}
	= 0, \label{conte}\\
	\frac{1}{n_{\mu}} \frac{1}{r^2}\frac{\partial}{\partial r}\left[r^2 n_{\mu} \left( \xi_{{b}r}^{1} - z_r^{1} \right)\right]
		+ \frac{1}{r\sin\theta}
	\left\{
		\frac{\partial}{\partial\theta} \left[\sin\theta \left(\frac{\partial Q^{1}_b}{\partial \theta} - \frac{\partial Q_z^{1}}{\partial \theta} \right)\right]
		- \frac{m^2 (Q^{1}_b-Q_z^{1})}{\sin\theta}
	\right\}
	= 0 \label{contmu}
\end{eqnarray}
in a stratified star [when $d(n_e/n_\mu)/dr \neq 0$] imply 
\begin{eqnarray}
\xi_{{b}r}^{1} - z_r^{1}=0, \label{radrestr} \\
	Q^{1}_b -Q_z^{1}=0. \label{anglerestr}
\end{eqnarray}
Expressing 
$Q^{1}_{b\;m+1,m}$ and $Q^{1}_{z\; m+1,m}$ 
in Eq.\ (\ref{anglerestr}) 
through, respectively,
$T^{0}_{b\;mm}$, $\xi_{{b}r, {m+1,m}}^{1}$ and
$T^{0}_{z\; mm}$, $z_{r, {m+1,m}}^{1}$, 
and using Eq.\ (\ref{radrestr}), we find
\begin{eqnarray}
T^{0}_{z\; mm}=\frac{(1 + m) \sigma_{1} T^{0}_{b\;mm}}{(1+m)\sigma_{1}  + 2 \Delta h}. \label{Trelation}
\end{eqnarray}
The $\phi$-component of the Euler equation  allows one to express 
$\delta P^{0}_{m+1,m}$ through $T^{0}_{b\;mm}$, 
while the $\phi$-component of the SF equation 
can be used to
present 
$\delta \eta^{0}_{\mu\;m+1,m}$
as a function of $\delta \eta^{0}_{e\;m+1,m}$ and $T^{0}_{z\;mm}$.

Substituting the obtained expressions for $\delta P^{0}_{m+1,m}$ and $\delta \eta^{0}_{\mu\;m+1,m}$ into the $r$-component of the Euler equation, we derive an ODE of the form:
\begin{eqnarray}
	\d{}{r} T_{b\;mm}^{0}
	- \frac{m}{r} T_{b\;mm}^{0}
	+ b_1(r) T_{z\;mm}^{0}
	+ c_1(r) \delta \eta^{0}_{e\;m+1,m}
	= 0  \label{rEuler}
,
\end{eqnarray}
while substitution of these expressions into the $r$-component of the SF equation gives
\begin{eqnarray}
	\d{}{r} \left(\frac{T_{z\;mm}^{0} h_1}{n_\mu}\right)
	- \frac{m}{r} \frac{T_{z\;mm}^{0} h_1}{n_\mu}
	+ c_2(r) \delta \eta^{0}_{e\;m+1,m}
	= 0  \label{rsfl}
.
\end{eqnarray}
Equations (\ref{Trelation}), (\ref{rEuler}), and (\ref{rsfl}) allow us to express $\delta \eta^{0}_{e\;m+1,m}$ through $T_{b\;mm}^{0}$, 
which results in the equation
\begin{eqnarray}
	\d{}{r} T_{b\;mm}^{0}- \frac{m}{r} T_{b\;mm}^{0}
	-\sigma_{1} a_1(r,\sigma_{1},\Delta h) T_{b\;mm}^{0}
	= 0   \label{Teq}
.
\end{eqnarray}
The functions $b_1(r)$, $c_1(r)$, $c_2(r)$, and $a_1(r,\sigma_{1},\Delta h)$ in Eqs.\ (\ref{rEuler})--(\ref{Teq})
are known (have been found), 
but their actual form is not important for us here.

Equations (\ref{xibeq}) and (\ref{Teq}) describe the r-mode oscillations in the leading order in rotation frequency 
in SF $npe\mu$ NS core up to the first-to-the-leading-order correction in the entrainment under the assumption that expansions (\ref{exp_start})--(\ref{exp_end}) are valid. 
This system, Eqs.\ (\ref{xibeq}) and (\ref{Teq}), 
should be supplemented with 
a number of boundary conditions. 
Regularity in stellar center (at $r\rightarrow 0$) requires 
\begin{eqnarray}
T_{b\;mm}^{0}\propto r^m,\\
\xi_{{b} r, m+1,m}^{1} = \frac{(1 + m)^2 \sigma_{1}}{2 (1 + 2 m)} T_{b\;mm}^{0}. \label{bccenter}
\end{eqnarray}

Since the particle number densities in our background model are continuous,  
the continuity equations imply continuity of baryon and lepton radial displacements, 
$\xi_{{b} r, m+1,m}^{1}$ and $\xi_{ r, m+1,m}^{1}$. 
This condition leads to the requirement of vanishing $z_{r, m+1,m}^{1}$ at the SF interface.
Moreover, $r$ and $\phi$ components of the Euler and SF equations require continuity of the functions $T_{b\;mm}^{0}$ and $T_{z\;mm}^{0}$ throughout the star.

At the stellar surface (where we assume that the matter is non-SF and barotropic) 
we require the Lagrangian perturbation of the pressure to be zero, $\Delta P=0$, which means, in the leading order in rotation,
\begin{eqnarray}
\xi_{{b} r, m+1,m}^{1}(R) =0. \label{bcsurf}
\end{eqnarray}

Consider now, for simplicity, 
a two-layer star composed of SF $npe\mu$ core and a barotropic single-fluid crust.
To find the eigenfunctions in the whole star 
we have to employ oscillation equations in the crust as well:
\begin{eqnarray}
\d{}{r} T_{b\;mm}^{0}
	- \frac{m}{r} T_{b\;mm}^{0}=0, \label{Tcrust}\\
	\d{}{r} \xi_{{b} r, m+1,m}^{1}
	+\left(\frac{n_{b}'}{n_{b}}+\frac{3+m}{r}\right) \xi_{{b} r, m+1,m}^{1}
	- \frac{(1 + m)^2 (3 + 2 m) \sigma_{1}}{(2+4m)r}T^{0}_{b\;mm}
	= 0.  \label{xibcrust}
\end{eqnarray}

Assume first that $\sigma_{1}=0$ (vanishing $\sigma_{1}$ was a solution for normal r-mode in $npe$ NSs, 
see Ref.\ \cite{dkg19}). 
Then equations in the core and in the crust coincide. 
Moreover, $\xi_{{b} r, m+1,m}^{1}=0$ throughout the star 
[due to Eqs.\ (\ref{xibeq}), (\ref{bccenter}), and (\ref{xibcrust})],
while 
in the core $z_{r, m+1,m}^{1}=\xi_{{b} r, m+1,m}^{1}=0$ and $T^{0}_{z\; mm}=0$ due to, respectively, 
Eqs.\ (\ref{radrestr}) and (\ref{Trelation}).
In addition, $T_{b\;mm}^{0} \propto r^m$. 
This solution meets all the boundary conditions and describes the normal nodeless r-mode.

Now, if $\sigma_{1}\neq 0$, then, 
integrating Eqs.\ (\ref{xibeq}), (\ref{Teq}), and (\ref{Tcrust}), 
(\ref{xibcrust}), we have to meet three boundary conditions: in the stellar center (\ref{bccenter}),
at the surface (\ref{bcsurf}), and at the core-crust interface: 
$\xi_{{b} r, m+1,m}^{1}=z_{r, m+1,m}^{1}=0$. 
At the same time, 
we have only {\it two} integration constants 
(one of which defines the oscillation amplitude) 
and undefined value of $\sigma_{1}$. 
This is clearly not enough to meet all the boundary conditions;
our system appears to be overdetermined.
This happens because of restrictions (\ref{radrestr}) and (\ref{Trelation}). 
We come to conclusion
that no oscillations at non-zero entrainment are possible 
with $\sigma$ vanishing at $\Omega \rightarrow 0$, 
except for the normal r-mode. 
However, this conclusion looks to be unphysical, 
since it is hard to imagine
that account for
even infinitely small entrainment could eliminate
SF modes with $\sigma$ vanishing at $\Omega \rightarrow 0$.

\subsection{Successful scheme}
\label{}

To demonstrate that 
such modes do exist,
below we do not restrict ourselves to the leading order in rotation frequency, 
but instead account for the next-to-the-leading-order 
corrections in rotation and in the entrainment {\it simultaneously}. 
Such an approach allows us to relax the restrictions (\ref{radrestr}) and (\ref{Trelation}), 
and find a solution for SF r-mode oscillations at non-vanishing entrainment. 
In what follows we adopt the following expansions:
\begin{eqnarray}
	\sigma 
	= (\sigma_{0}  + \sigma_{1})\Omega 
	= \left(\frac{2}{m+1} + \sigma_{1}\right)\Omega  
	\label{expan1}
,\\
	\xi_{{b} r}= \xi_{{b} r}^{1}  
,\quad
	T_b = T^{0}_b + T^{1}_b   
,\quad
	Q_b = Q^{1}_b  
,\quad
	z_r = z_r^{1}  
,\quad
	T_z = T_z^{0} + T_z^{1}  
,\quad
	Q_z = Q_z^{1} 
	\label{expan2}
.
\end{eqnarray}
The leading order in both rotation and entrainment of each quantity is labeled with the index $0$, while 
index $1$ denotes next to the leading order corrections, both in entrainment and rotation. 
For example, as we shall see below, 
$\sigma_{1}$ behaves as $\Omega^2$ 
at 
high rotation frequency,
and does not depend on the rotation frequency at small rotation rate and finite entrainment. 
Since in the absence of entrainment the $r$-modes are purely toroidal 
in the leading order in rotation,
the radial and poloidal displacements in the zeroth order vanish,
$\xi_{{b} r}^{0} = z_r^{0} = Q^{0}_b = Q_z^{0} = 0$.
Equations describing the leading order are the same as in Sec.\ \ref{subs:failed} 
with the solution for the eigenfrequency and eigenfunctions given by Eqs.\ (\ref{sigma0})--(\ref{T}).

To find $\sigma_{1}$ and the functions $T_{b\;mm}^{0} (r)$ and $T_{z\;mm}^{0} (r)$,
we again consider the continuity equations \eqref{cont b}--\eqref{cont l},
as well as $r$, $\phi$, and $\theta$-components of the Euler equation \eqref{euler} and the SF equation \eqref{sfl1}.
As in Sec.\ \ref{subs:failed}, 
the $\theta$-components of the Euler and SF equations 
allow us to express $Q_{b\;m+1,m}^{1} (r)$ through $\xi_{{b} r\, m+1,m}^{1} (r)$ 
and $T_{b\;mm}^{0} (r)$, and $Q_{z \,m+1,m}^{1} (r)$ 
through $z_{r\, m+1,m}^{1} (r)$ and $T_{z \,mm}^{0} (r)$. 
The relations, however, slightly differ from those in Sec.\ \ref{subs:failed}, 
because here we account for the oblateness of an NS 
(since it is $\propto \Omega^2$), 
as it is described in Ref.\ \cite{kg17}. 
Oblateness gives rise to additional terms 
in the $\theta$-components of the Euler and SF equations, 
as well as in the continuity equations, 
but does not affect our approach qualitatively.
The $r$ and $\phi$-components of the Euler and SF equations remain unaffected and 
give us again [see Eqs.\ (\ref{rEuler})--(\ref{rsfl})]
\begin{eqnarray}
	\d{}{r} T_{b\;mm}^{0}
	- \frac{m}{r} T_{b\;mm}^{0}
	+ b_1(r) T_{z\;mm}^{0}
	+ \frac{c_1(r)}{\Omega^2} \delta \eta_{e\;m+1,m}
	= 0,\label{EulerS}\\
	\d{}{r} \left(\frac{T_{z\;mm}^{0} h_1}{n_\mu}\right)
	- \frac{m}{r} \frac{T_{z\;mm}^{0} h_1}{n_\mu}
	+ \frac{c_2(r)}{\Omega^2} \delta \eta_{e\;m+1,m}
	= 0.  \label{sflS}
\end{eqnarray}

The main difference from Sec.\ \ref{subs:failed} is the continuity equations. 
Now, in contrast to Sec.\ \ref{subs:failed}, 
they contain the number density perturbations.
As a result,
leptonic continuity equations
remain nondegenerate
[i.e., the constraints (\ref{radrestr}) and (\ref{anglerestr}) should not be satisfied].
The coefficients at the polynomial $P_{m+1}^m$ in the three continuity equations 
(for baryons, electrons and muons), as well as Eqs.\ (\ref{EulerS})--(\ref{sflS}) 
allow us to express an unknown function 
$\delta \eta_{e\;m+1,m}$,
and eventually arrive at
the following system of 
oscillation equations:
\begin{eqnarray}
	\d{}{r} T_{b\;mm}^{0}
	-A_1(r) T_{b\;mm}^{0}
	-B_1(r) T_{z\;mm}^{0}
	-\frac{C_1(r)}{\Omega^2} (\xi_{{b} r, m+1,m}^{1}-z_{r, m+1,m}^{1})
	= 0,   \label{Teq1}\\
\d{}{r} {T_{z\;mm}^{0}}
	-A_2(r) T_{b\;mm}^{0}
	-B_2(r) T_{z\;mm}^{0}
	-\frac{C_2(r)}{\Omega^2} (\xi_{{b} r, m+1,m}^{1}-z_{r, m+1,m}^{1})
		= 0,  \label{Tzeq1}\\
		\d{}{r} {\xi_{{b} r, m+1,m}^{1}}
	-\left[\sigma_1 \frac{(1 + m)^2 (3 + 2 m)}{(2+4m)r}+\Omega^2 A_{3}(r)\right]	T^{0}_{b\;mm}
	-\Omega^2 B_{3}(r)	T^{0}_{z\;mm}\nonumber \\
	-C_3(r)\xi_{{b} r, m+1,m}^{1}
-D_3(r)z_{r, m+1,m}^{1}=0, \label{xibeq1}\\
	\d{}{r} z_{r, m+1,m}^{1}
	-\Omega^2 A_{4}(r)	T^{0}_{b\;mm}
	-\left[\frac{(1 + m) (3 + 2 m)(\sigma_1+m \sigma_1+2\Delta h)}{(2+4m)r}+\Omega^2 B_{4}(r)\right]	T^{0}_{z\;mm}\nonumber \\
	-C_4(r)\xi_{{b} r, m+1,m}^{1}
-D_4(r)z_{r, m+1,m}^{1}=0, \label{zeq1}
\end{eqnarray}
where $A_i(r)$, $B_i(r)$, $C_i(r)$, $D_i(r)$ ($i=1,2,3,4$) are some known functions of the 
radial coordinate.
Regularity of these equations at the stellar center ($r\rightarrow 0$) implies
\begin{eqnarray}
T^0_{b\;mm},T^0_{z\;mm},\xi_{{b} r, m+1,m}^{1},z_{r, m+1,m}^{1}\propto r^m, \label{bccenter10}\\
\xi_{{b} r, m+1,m}^{1}=\frac{(1+m)^2}{2+4m}\sigma_1 T^0_{b\;mm}, \label{bccenter11}\\
z_{r, m+1,m}^{1}=\frac{1+m}{2+4m} (\sigma_1 + m \sigma_1 + 2 \Delta h)T^0_{z\;mm}. \label{bccenter12}
\end{eqnarray}
Vanishing Lagrangian perturbation of the pressure at the surface leads to
the condition
\begin{eqnarray}
T^0_{b\;mm}(R)=\frac{(1 + m)^2 (1 + 2 m) \xi_{{b} r, m+1,m}^{1}(R)\; P_0'(R)}{4 m \Omega^2\; w_0(R)\; R}. \label{bcsurf1}
\end{eqnarray}
One also has to require vanishing of $z_{r, m+1,m}^{1}$ at the SF interface 
and continuity of the functions $\xi_{{b} r, m+1,m}^{1}$, $T^0_{b\;mm}$, and $T^0_{z\;mm}$ throughout the star. 

Consider again, for simplicity, a two-layer star, 
consisting of the SF $npe\mu$ core and a barotropic single-fluid crust.
Having six integration constants (two in the crust and four in the core), 
we can meet all the necessary boundary conditions 
[(\ref{bccenter11}), (\ref{bccenter12}), (\ref{bcsurf1}), 
continuity of $\xi_{{b} r, m+1,m}^{1}$ and $T^0_{b\;mm}$, 
and vanishing of $z_{r, m+1,m}^{1}$ at the core-crust interface), 
by
adjusting the value of $\sigma_1$. 
Thus, the approach developed  
in this section
allows us to 
restore the missing SF rotational modes in the spectrum.

\subsection{Limit of vanishing rotation rate}
\label{zerolimit}
Let us now analyze the solution to the system (\ref{Teq1})--(\ref{zeq1})
in the limit $\Omega \rightarrow 0$,
which got us into trouble in Sec.\ \ref{subs:failed}.
If $T^0_{z\;mm}=0$, then $\Delta h$ does not enter the equations and
we have the standard ordering, 
$\sigma_1 \sim \Omega^2$, $\xi_{{b} r, m+1,m}^{1} \sim z_{r, m+1,m}^{1} \sim \Omega^2 T^0_{b\;mm}$,
$\delta P\sim \delta \eta_{e}\sim\delta \eta_{\mu} \sim \Omega^2$.
This is the normal r-mode.
For SF modes $T^0_{z\;mm}\neq 0$ and the standard ordering is not valid 
[see the term $\propto T^0_{z\;mm}$ in Eq.\ (\ref{zeq1})].
Instead, as we demonstrate below, 
$\sigma_1$ appears to be finite and the ordering 
of eigenfunctions is the following:
$\xi_{{b} r, m+1,m}^{1} \sim z_{r, m+1,m}^{1} \sim \Omega T^0_{b\;mm} \sim \Omega T^0_{z\;mm}\sim \Omega^2 dT^0_{b\;mm}/dr \sim \Omega^2 dT^0_{z\;mm}/dr \sim \Omega\, d\xi_{{b} r, m+1,m}^{1}/dr \sim \Omega\, dz_{r, m+1,m}^{1}/dr\sim  \delta \eta_{e}\sim \delta \eta_{\mu} \sim \delta P/\Omega$.

To start with, we rewrite Eqs.\ (\ref{Teq1})--(\ref{zeq1}) 
assuming the ordering above 
and neglecting the terms, which are small according to this ordering.
Doing this, we step away from the stellar center, in order not to deal with the additional small parameter, $r$
%
\footnote{In our analysis we also assume that neutrons are SF, $T<T_{{\rm c}n}$,
everywhere in the $npe\mu$ core. 
This assumption allows us to avoid the
singularity related to infinite growth of the function $h_1$ 
at the boundary of the SF region, at $T\rightarrow T_{{\rm c}n}$. 
Note that 
$h_1\rightarrow \infty$
at $T\rightarrow T_{{\rm c}n}$, 
and the excluded terms can be large. 
However, 
numerical calculations
show that the consideration below is relevant at any temperature.}:
%
\begin{eqnarray}
	\d{}{r} T_{b\;mm}^{0}
	-\frac{C_1(r)}{\Omega^2} (\xi_{{b} r, m+1,m}^{1}-z_{r, m+1,m}^{1})
	= 0,  \label{eq1} \\
\d{}{r} {T_{z\;mm}^{0}}
	-\frac{C_2(r)}{\Omega^2} (\xi_{{b} r, m+1,m}^{1}-z_{r, m+1,m}^{1})
		= 0,  \label{eq2}\\
		\d{}{r} {\xi_{{b} r, m+1,m}^{1}}
	-\sigma_1 \frac{(1 + m)^2 (3 + 2 m)}{(2+4m)r}	T^{0}_{b\;mm}
	=0, \label{eq3}\\
	\d{}{r} z_{r, m+1,m}^{1}
		-\frac{(1 + m) (3 + 2 m)(\sigma_1+m \sigma_1+2\Delta h)}{(2+4m)r}	T^{0}_{z\;mm}=0. \label{eq4}
\end{eqnarray}
The above equations
can be combined to (again, up to the terms, leading in $\Omega$)
\begin{eqnarray}
	\d{T}{r} -\frac{K(r,\sigma_1)}{\Omega^2} \xi
	= 0,  \label{eq11} \\
		\d{\xi}{r}	-T	=0, \label{eq111}
\end{eqnarray}
where we define
\begin{eqnarray}
\xi\equiv {\xi_{{b} r, m+1,m}^{1}}-z_{r, m+1,m}^{1},\label{xidef}
\\
T \equiv \sigma_1 \frac{(1 + m)^2 (3 + 2 m)}{(2+4m)r}	T^{0}_{b\;mm}-\frac{(1 + m) (3 + 2 m)(\sigma_1+m \sigma_1+2\Delta h)}{(2+4m)r}	T^{0}_{z\;mm},\label{Tdef}\\
K(r,\sigma_1)\equiv \sigma_1 \frac{(1 + m)^2 (3 + 2 m)}{(2+4m)r}C_1(r)-\frac{(1 + m) (3 + 2 m)(\sigma_1+m \sigma_1+2\Delta h)}{(2+4m)r}C_2(r).
\end{eqnarray}  
Writing down the ratio of Eqs.\ (\ref{eq11}) and (\ref{eq111}), 
\begin{equation}
T dT=\frac{K(r,\sigma_1)}{\Omega^2}\xi d \xi,
\end{equation} 
we find that $\xi \sim \Omega T$, 
or $\xi_{{b} r, m+1,m}^{1}\sim z_{r, m+1,m}^{1}\sim \Omega T^{0}_{b\;mm}\sim \Omega T^{0}_{z\;mm}$ 
[from Eqs.\ (\ref{eq1})--(\ref{eq2}) 
it follows that $T^{0}_{b\;mm}\sim	T^{0}_{z\;mm}$, 
while Eqs.\ (\ref{eq3})--(\ref{eq4}) imply that $\xi_{{b} r, m+1,m}^{1}\sim z_{r, m+1,m}^{1}$]. 
Notably, this ordering is different from the standard one, 
which takes place at vanishing entrainment, 
when we account for the rotational corrections only. 
In that case $\Delta h=0$, $\sigma_1 \propto \Omega^2$ and one finds from Eqs.\ (\ref{Teq1})--(\ref{zeq1})
$\xi_{{b} r, m+1,m}^{1}\sim z_{r, m+1,m}^{1}\sim \Omega^2 T^{0}_{b\;mm}\sim \Omega^2 T^{0}_{z\;mm}$.

On the other hand, Eqs.\ (\ref{eq11})--(\ref{eq111}) can be rewritten as
\begin{eqnarray}
	\frac{d^2 \xi}{dr^2}
	-\frac{K(r,\sigma_1)}{\Omega^2} \xi
	= 0.
	\label{eq22}
\end{eqnarray}
If $K(r,\sigma_1)<0$ in some range of $r$, we have an oscillating 
solution there
(with a vanishing wavelength at $\Omega \rightarrow 0$), 
whereas
in the region with $K(r,\sigma_1)>0$
the eigenfunctions exhibit exponential behavior
(with an infinite exponent at $\Omega \rightarrow 0$).
This means that, if we are interested in the solution with finite number of nodes, 
we cannot have extended regions with $K(r,\sigma_1)<0$ 
in the limit $\Omega \rightarrow 0$,
since they contain an infinite number of nodes.
At the same time, we also cannot have $K(r,\sigma_1)>0$ everywhere, 
since this would not allow us to 
meet all the boundary conditions
due to different ordering of eigenfunctions in the SF $npe\mu$ core and in the remaining star. 
The solution is only possible if $K(r,\sigma_1)>0$ everywhere except for one point 
(infinitely small region at $\Omega\rightarrow 0$), 
where $K(r,\sigma_1)$ vanishes (see Appendix for details).

As a result, 
$\sigma_1$ for SF modes tends to the finite value at $\Omega \rightarrow 0$, 
which can be found from the condition 
\begin{equation}
{\rm min}[K(r,\sigma_1)]=0,   \label{condOmega0}
\end{equation} 
i.e., the minimum of the function $K(r,\sigma_1)$ 
in the SF $npe\mu$-core must vanish. 
Moreover, all the overtones must have the same $\sigma_1$ 
[since infinitesimal variation of $\sigma_1$ 
allows one to increase the number of nodes in the infinitely small region, where $K(r,\sigma_1)<0$]. 
The function $K(r,\sigma_1)$ may have the minimum at any point in SF
$npe\mu$ core,
in particular, in the centre 
or at the muon onset density. 
For example, in the zero-temperature limit 
the minimum occurs 
at the outer boundary of the $npe\mu$ core for APR EOS (see Sec.\ \ref{sec:input}), 
while for BSk24 EOS it is located at the stellar center.

It is now interesting to discuss why our attempt to find the solution in Sec.\ \ref{subs:failed}
was not successful.
First of all, in Sec.\ \ref{subs:failed} 
we assumed the standard ordering of eigenfunctions, 
which is not valid for SF modes, 
as we demonstrated in this section. 
While the $\phi$-component of the Euler equation 
implies 
$\delta P\sim \Omega^2 T^0_b$, 
the imbalances of chemical potentials 
scale as $\delta \eta_l \sim \Omega T^0_b$ 
due to the scaling $\xi_{{b} r}^{1} \sim z_{r}^{1} \sim \Omega T^0_b$. 
As a result, perturbations of particle number densities scale as 
$\delta n_i \sim \Omega T^0_b$. 
At first glance, it seems that
with this ordering  
we can skip $\delta n_l$ in Eqs.\ (\ref{conte}) and (\ref{contmu}) 
in the leading order in $\Omega$, because $d\xi_{{b}r}^{1}/dr \sim T^0_b$. 
However, if we consider the difference of 
Eqs.\ (\ref{conte}) and (\ref{contmu}), 
$d\xi_{{b}r}^{1}/dr$ and the functions $Q^1_b$ and $Q^1_z$ will cancel out, leaving us with
\begin{equation}
\delta n_e-\delta n_\mu+\left( \xi_{{b}r}^{1} - z_r^{1} \right)\left(\frac{1}{n_{e}} \frac{\partial n_{e} }{\partial r} -\frac{1}{n_{\mu}} \frac{\partial n_{\mu} }{\partial r}\right)=0.
\end{equation}
All the terms in this equation are of the same order in $\Omega$. 
Thus, the constraints (\ref{radrestr}) and (\ref{anglerestr}) of Sec.\ \ref{subs:failed} are not applicable
in our situation.

\section{Physics input}
\label{sec:input}

In our numerical calculations we adopt a model of a three-layer star, consisting of the crust, outer core (composed of neutrons, protons, and electrons), and the inner core (containing, in addition, muons). 
Since the crust does not affect eigenfrequencies of global NS oscillations strongly, 
it is treated within a simplest model of
barotropic one-component fluid. The core is modeled more realistically, adopting modern equations of state (EOSs), which lead to non-barotropic behavior of the core liquid due to composition gradients, and accounting for possible superfluidity/superconductivity of baryons. 

We consider two EOSs in the core. 
The first one is essentially the same as in Ref.\ \cite{kg17}. 
It uses parametrization \cite{hh99} of Akmal-Pandharipande-Ravenhall (APR) EOS \cite{apr98}, 
and adopts entrainment matrix from 
Ref.\ \cite{gh05}. 
The second EOS is constructed with the BSk24 energy-density functional (BSk24 EOS) \cite{gcp13,pcp18}. 
The entrainment matrix for this EOS is calculated self-consistently,
by extracting nucleon Landau parameters from this functional and then following the prescription of 
Refs.\ \cite{gkh09a,gkh09b,ghk14} (see Ref.\ \cite{kg21} for details).

\begin{figure}
	 \begin{center}
	  \leavevmode	
	\includegraphics[width=7in]{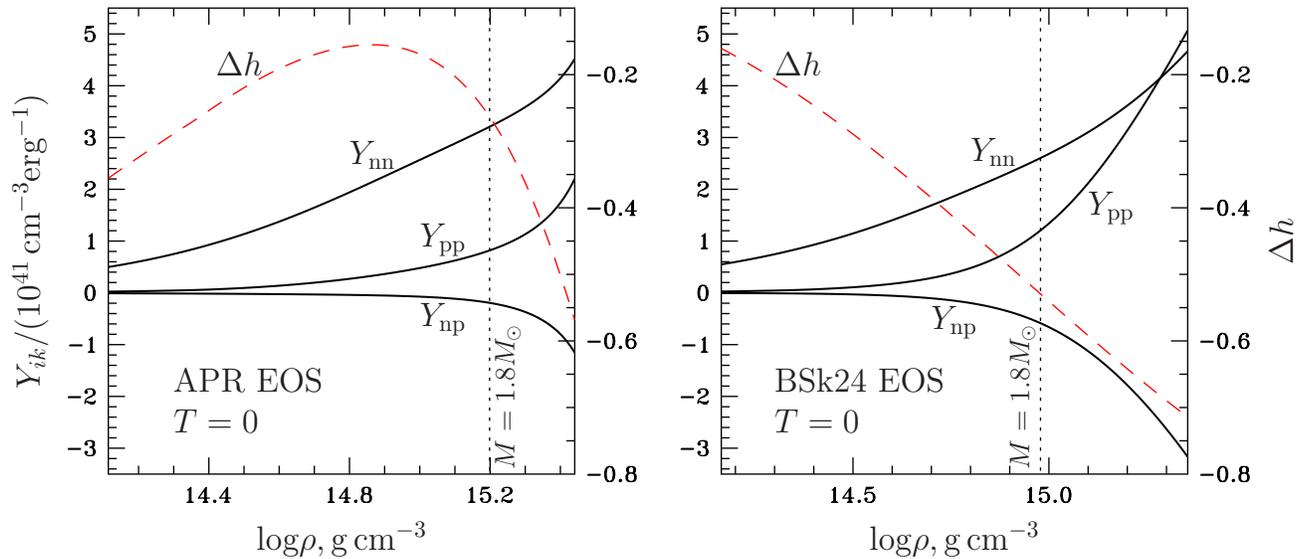}
	 \end{center}
 \caption{Entrainment matrix elements $Y_{ik}$ and 
 	the parameter $\Delta h$ versus density in the zero-temperature limit for APR EOS (left panel) and BSk24 EOS (right panel).
}
	\label{Fig:Yik} 
\end{figure}

To illustrate the behavior of the entrainment matrix 
we plot its elements $Y_{ik}$ ($i,k=n,p$) as functions of density 
for two EOSs 
in the limit of vanishing temperature
(see solid lines in the left and right panels of Fig.\ \ref{Fig:Yik}). 
In both panels density ranges from its value at the core-crust interface 
up to the central density in the limiting configuration of an NS with the maximum mass. 
Vertical dots show central densities of an NS with $M=1.8M_\odot$.
In the same plot we also present by dashes the 
parameter $\Delta h$. 
One can see that  $\Delta h$ can hardly be 
considered small
at high densities. However, since the r-mode eigenfunctions
are mostly localized in the outer core,
in what follows we 
shall treat
$\Delta h$ as a small parameter and use expansions discussed in the previous section. 
Note that the method of expansion in the entrainment has been proved to be rather accurate in Ref.\ \cite{dkg19}, where the rotational spectrum 
of an NS with the superfluid $npe$ core was studied. 
In that reference we calculated the (temperature-dependent) 
eigenfrequencies of superfluid r-mode by two different methods,
either using the perturbation theory in the entrainment or not using it (exact calculation).
Both approaches give very similar temperature-dependent spectra 
(compare upper solid red line and dot-dashed line in figure 2 of Ref.\ \cite{dkg19}).
%

\begin{figure}
    \begin{center}
        \leavevmode
        \includegraphics[width=6in]{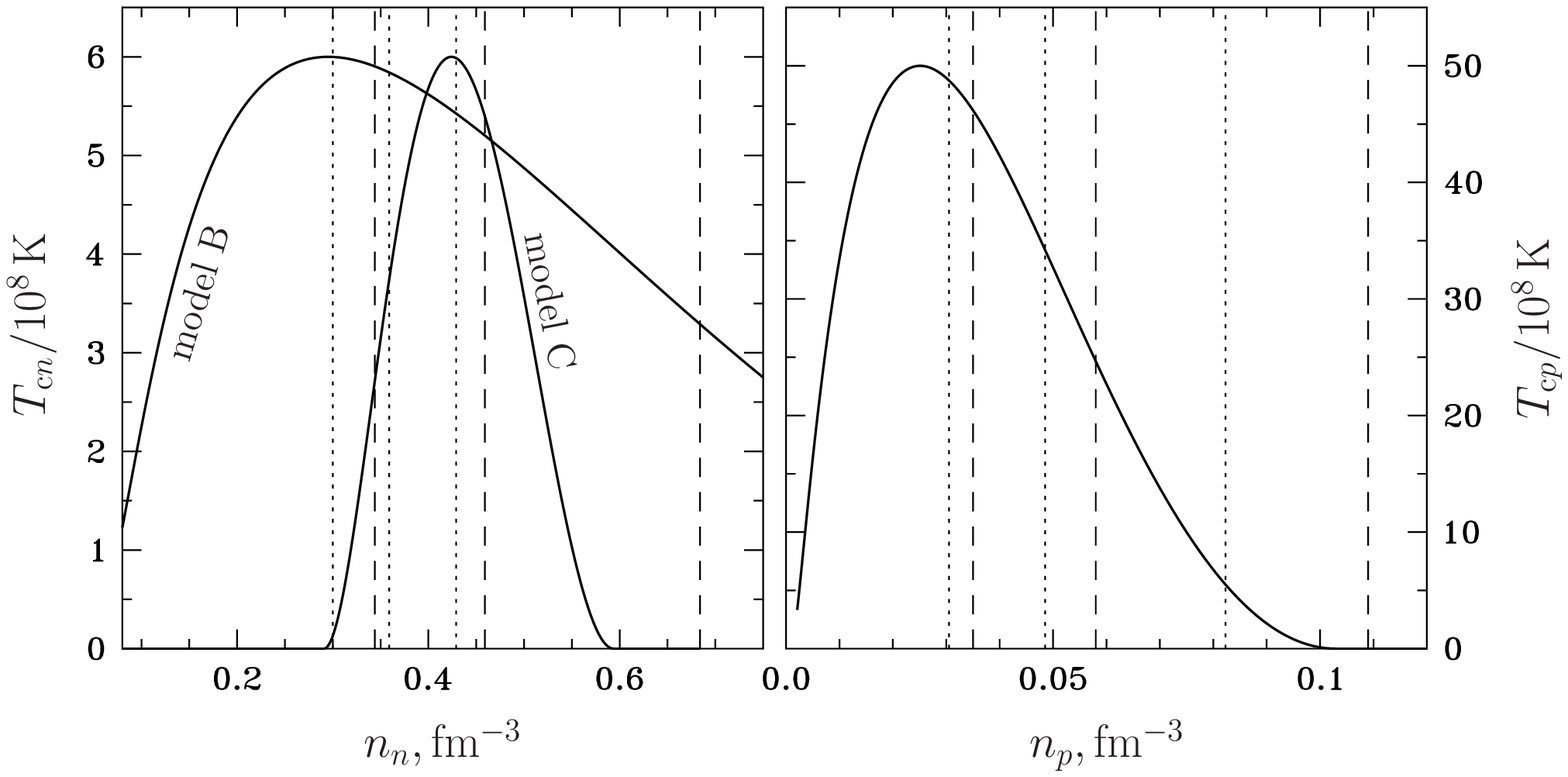}
    \end{center}
    \caption{Left panel: Neutron (local) critical temperatures $T_{{\rm c}n}$ versus $n_n$ for models B and C. 
      Right panel: Proton (local) critical temperature $T_{{\rm c}p}(n_p)$ (the same for models B and C).
Vertical lines (dashes for APR EOS, dots for BSk24 EOS) indicate central number densities of (from left to right) $M=1.0M_\odot$, $M=1.4M_\odot$, and $M=1.8M_\odot$ NSs. 
		}
    \label{Fig:Tc}
\end{figure}

For both EOSs we consider three SF models. 
In the first one neutron and proton local critical temperatures of SF onset 
are density independent (model A) 
and equal $T_{{\rm c}n}=6\times 10^8\,\rm K$ and 
$T_{{\rm c}p}=5\times 10^9\,\rm K$ for neutrons and protons, respectively. 
Although this model, as it is, is not realistic from point of view of microscopic calculations, 
it can be viewed
as a limiting case for realistic wide critical temperature profiles. 
In our second and third models (models B and C) we adopt density-dependent (local) critical temperature profiles,
$T_{{\rm c}n}(n_n)$ and $T_{{\rm c}p}(n_p)$
(see Fig.\ \ref{Fig:Tc}).
These models coincide with the models I and II from Ref.\ \cite{kgd20}.
The
$T_{{\rm c}p}$ profile is the same for both models, 
while $T_{{\rm c}n}$ profiles differ.
Model B has a wide $T_{{\rm c}n}$ profile, 
similar to that predicted by modern microscopic calculations (see, e.g., \cite{drddwcp16,sc19}).
Model C, on the contrary, 
describes a narrow $T_{{\rm c}n}$ profile, 
which leaves
the outer core non-SF at 
any reasonable temperature.
Similar (purely phenomenological)
profiles have been used in a number of works 
(e.g., \cite{gkyg04,gkyg05,syhhp11,CasA13})
to successfully explain thermal properties 
of isolated neutron stars within the minimal cooling scenario \cite{plps04,gkyg04}.
All the adopted profiles have maximum critical temperatures that do not contradict 
both the existing data on cooling NSs \cite{gkyg04,plps04,gkyg05,syhhp11,page11,CasA13,CasA15,bhsp18}
and microscopic calculations \cite{ls01, yls99,gps14,dlz14, drddwcp16, sc19}.

\section{Results}
\label{sec:res}

\begin{figure}
    \begin{center}
        \leavevmode
        \includegraphics[width=7in]{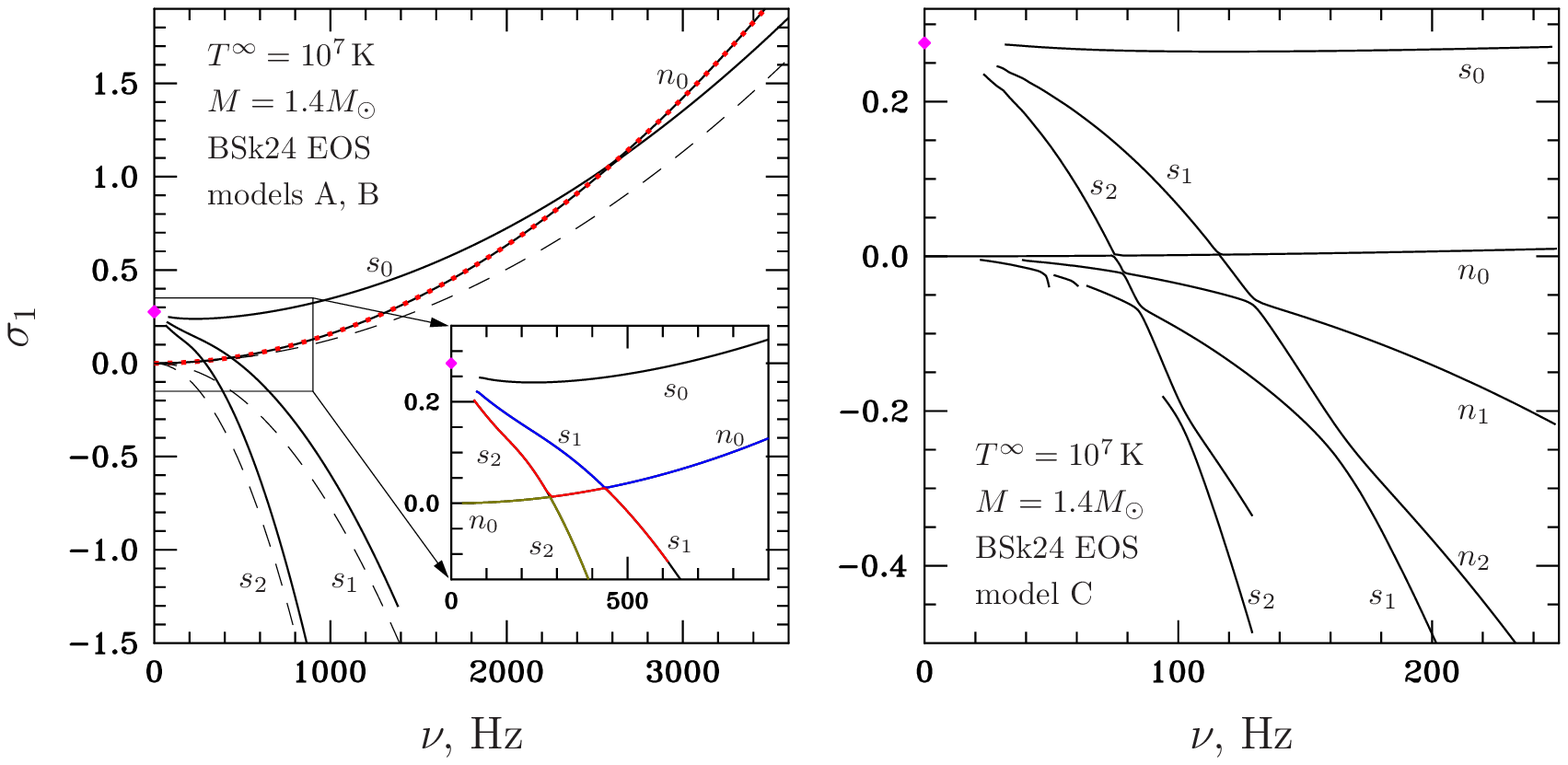}
    \end{center}
    \caption{$\sigma_1$ versus $\nu$ at $T^\infty=10^7 \,\rm K$ for models A and B (left panel) and model C (right panel).
    	The spectrum is plotted for an NS with $M=1.4 M_\odot$ assuming BSk24 EOS. See text for details.
		}
    \label{Fig:sigma1Omega}
\end{figure}

We present the results for the $l=m=2$ r-modes, 
since $l=m=2$ normal r-mode without nodes is known to be the most unstable one \cite{ak01}.
Fig. \ref{Fig:sigma1Omega} shows how the correction $\sigma_1$ depends on the rotation frequency $\nu$
%
\footnote{In the left panel the rotation frequency
ranges up to unphysical values of the order of $\nu \approx 3000\,\rm Hz$, 
which exceed the Kepler limit. 
Moreover, the low-frequency approximation is invalid at such 
frequencies:
We discuss them here only for the sake of completeness, 
to give an impression to the reader, 
where the avoided-crossing of normal and SF nodeless r-modes generally take place.}
%
for an NS with $M=1.4M_\odot$ 
and the stellar red-shifted
temperature $T^\infty=10^7\,\rm K$ (as seen by a distant observer). 
To plot the figure, we adopted BSk24 EOS 
and SF models from Sec.\ \ref{sec:input}. 
Left panel demonstrates the spectrum for models A and B. 
Since for these models $T_{{\rm c}n}^\infty \gg 10^7\,\rm K$ everywhere in the core
(for the chosen stellar configuration),
the spectrum 
for them
is indistinguishable.
For these models 
we find one normal nodeless r-mode ($n_0$), 
one SF nodeless r-mode ($s_0$), 
and an infinite set of SF modes with nodes (only first two overtones, $s_1$ and $s_2$, 
are presented in the figure).
Various oscillation modes are shown by solid lines; in the inset these lines are shown by different colors.
Note that the modes exhibit avoided-crossings with each other, 
altering their behavior from normal-like to SF-like and vice versa. 
Dots (red online) indicate normal nodeless r-mode 
(at different temperatures different r-modes behave as the normal one). 
The normal r-mode is virtually not affected by entrainment, 
while SF modes deviate strongly from their 
`vanishing-entrainment' 
behavior (shown by dashes), 
especially at small rotation frequencies. 
Diamond at $\nu=0$ shows the theoretically predicted limit 
of $\sigma_1$ at $\nu \rightarrow 0$ for SF modes,
defined by Eq.\ (\ref{condOmega0}).
One can see that the calculated curves tend to approach this limit 
(unfortunately, due to numerical issues we cannot carry out  our calculations at too small rotation frequencies). 
On the other hand, in the limit of rapid rotation, 
we see that $\sigma_1 \propto \Omega^2$, as expected.

Right panel shows the spectrum obtained for the model C. 
For this model, in addition to nodeless r-modes (normal and SF), 
we find both an infinite set of SF modes with nodes 
and an infinite set of normal modes with nodes. 
The plot shows the main harmonics ($n_0$ and $s_0$) 
and first two overtones ($n_1$, $n_2$, and $s_1$, $s_2$) of normal ($n$) and SF ($s$) modes. 
The modes exhibit avoided-crossings with each other. 
Irregularities 
of
the curves at low frequencies are due to the avoided-crossings with higher-order SF modes.
Again, the diamond at $\nu=0$ shows the theoretically predicted limit 
(defined by ${\rm min}\,[K(r,\sigma_1)]=0$
%
\footnote{For BSk24 EOS and $T^\infty=10^7\,\rm K$, 
$K(r)$ has a minimum at the stellar center for all 
SF models.},
%
see Eq.\ \ref{condOmega0}) for $\sigma_1$ at $\nu \rightarrow 0$ 
for SF modes.  
Our calculation demonstrates that the SF modes have a tendency to approach this limit,
while $\sigma_1$ for the normal modes tends to zero at $\nu \rightarrow 0$, 
as expected (see Sec.\ \ref{sec:r-mode}).

One can see that the spectra in both panels differ qualitatively.
The SF models A and B do not support normal modes with nodes, 
while model C supports them. 
This happens because for models A and B at $T^\infty=10^7\,\rm K$ 
there is no non-SF non-barotropic region in the star: 
the crust is assumed to be barotropic and the whole core is SF. 
At the same time, in model C the outer core is normal, and, 
since we use the non-barotropic EOS in the core, 
the outer core may support normal r-modes with nodes \cite{pbr81,yl00}.
In contrast to normal modes,
we have an infinite set of SF modes with nodes for all the three models, 
because for all of them there is a SF
region stratified by muons at $T^\infty=10^7\,\rm K$.

\begin{figure}
    \begin{center}
        \leavevmode
 \includegraphics[width=7in]{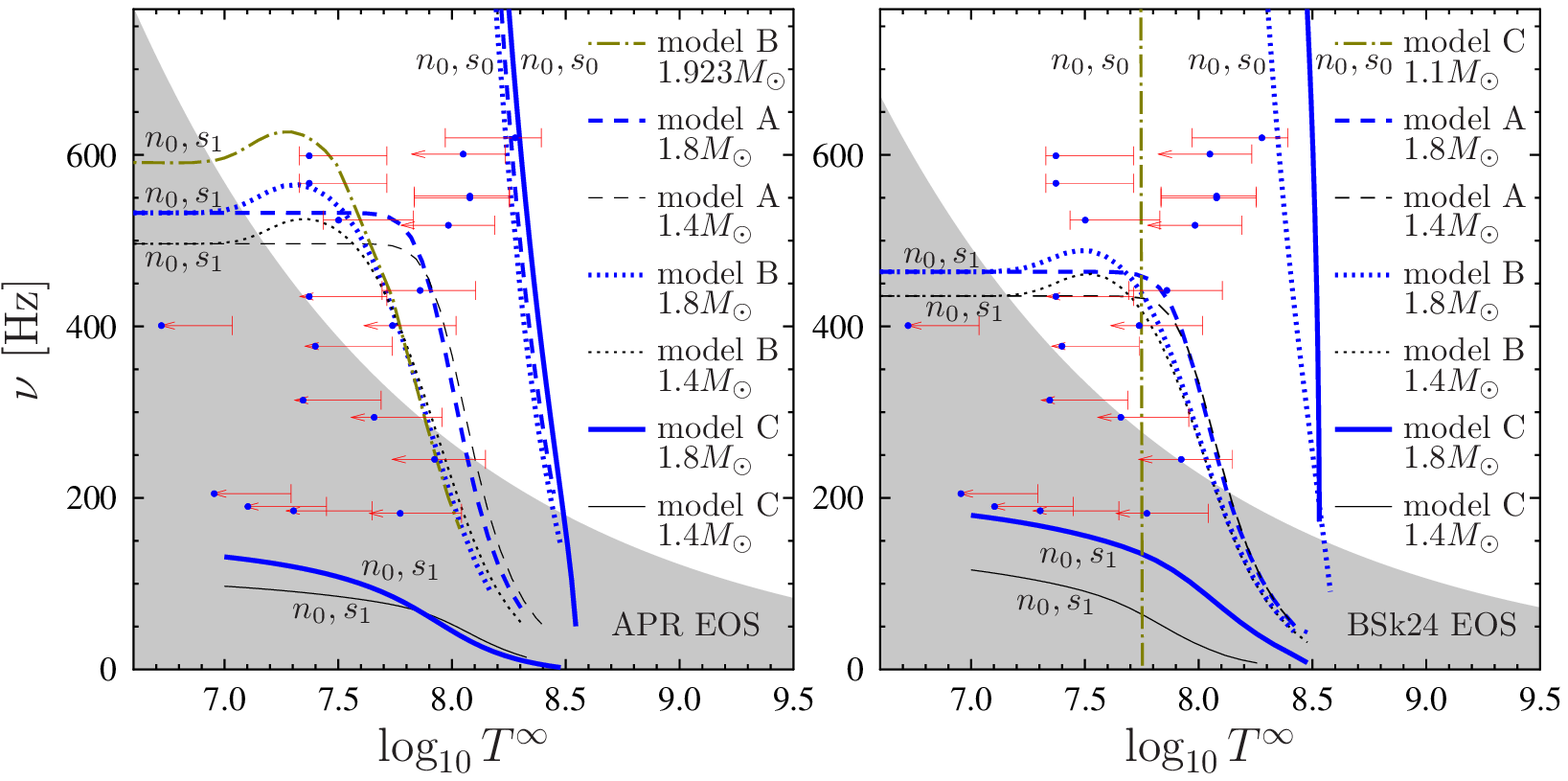}
    \end{center}
    \caption{Values of $T^\infty$ and $\nu$ 
    	at which normal r-mode experiences avoided-crossing with the main harmonic (marked with $n_0,s_0$) and first overtone (marked with $n_0,s_1$) of SF r-modes. 
    	Thick lines (dashes for the model A, dots for the model B, and solid lines for the model C) 
    	show results for an NS with $M=1.8M_\odot$, 
    	while thin lines (again dashes for the model A, dots for the model B, and solid lines for the model C) 
    	correspond to an NS with $M=1.4M_\odot$. 
    	Left panel: APR EOS, right panel: BSk24 EOS.
		In the vicinity of these curves normal r-mode experiences stabilizing interaction with corresponding SF r-mode (see Fig. \ref{Fig:peaks}). 
		Dot-dashed line in the left panel shows $\nu_{n_0,s_1}(T^\infty)$ 
		for the limiting configuration of 
		an NS ($M=1.923M_\odot$) in the model B. 
		In the right panel dot-dashed line shows $\nu_{n_0,s_0}(T^\infty)$ for $M=1.1M_\odot$ NS in the model C.
		Region filled gray is classical stable region defined by the shear viscosity only (calculated as described in Ref.\ \cite{gck14a} for an NS with $M=1.8M_\odot$). 
        Points with error bars describe available observational data 
        on NSs in LMXBs \cite{gck14b,pw17}. 
		}
    \label{Fig:inst}
\end{figure}

In Fig. \ref{Fig:sigma1Omega} 
one can see avoided-crossings of normal nodeless r-mode with SF r-modes.
At the rotation frequencies corresponding to the avoided-crossings, 
$\nu_{n_0,s_\alpha}$ ($\alpha=0,1,2,\ldots$), 
one should expect stabilization of normal r-mode 
by the resonance interaction with SF r-modes \cite{gck14a,gck14b}. 
At different temperatures 
the position $\nu_{n_0,s_\alpha}$ of the avoided crossings will be different.
Fig.\ \ref{Fig:inst} illustrates how $\nu_{n_0,s_\alpha}$ 
depend on $T^\infty$ for APR EOS (left panel) and BSk24 EOS (right panel).
Solid lines correspond to model C, 
dots represent model B, 
dashes correspond to the model A. 
Thick lines (both solid lines, dashes, and dots)
in both panels show the results for an NS with $M=1.8M_\odot$, 
thin lines -- for $M=1.4M_\odot$. 
Each curve is marked with the sign $n_0,s_0$ or $n_0,s_1$, 
which correspond to the avoided-crossings of normal nodeless r-mode with, 
respectively, the main harmonic or first overtone of SF r-modes. 
For comparison, points with error bars in Fig.\ \ref{Fig:inst} show the positions of the observed NSs in LMXBs 
taken from Refs.\ \cite{gck14b,pw17}.
Here we do not indicate the names of these sources to avoid cluttering of the figure, one can find the names in Fig.\ \ref{Fig:peaks}.

Generally (at not too high temperatures, see below), 
$n_0,s_0$ avoided crossing takes place at unphysically high rotation rates. 
Only when $T^\infty$ approaches the maximum value of $T_{{\rm c}n}^\infty$ 
in the region of the core, where neutron and proton SFs co-exist 
(denoted by $T_{{\rm c}n\, {\rm max}}^\infty$ in what follows), 
$\nu_{n_0,s_0}(T^\infty)$ decreases rapidly with increasing $T^\infty$ 
and falls to zero at $T^\infty=T_{\rm{cn\, max}}^\infty$. 
While the local value 
of $T_{{\rm c}n\, {\rm max}}$
equals
$T_{{\rm c}n\, {\rm max}}=6\times 10^8\,\rm K$ 
for
all our SF models, 
the red-shifted value, $T_{{\rm c}n\, {\rm max}}^\infty$, depends on the NS mass 
and EOS through the redshift parameter and varies in the range 
$T_{{\rm c}n\, {\rm max}}^\infty\sim (2.5-4)\times 10^8\,\rm K$. 
As a result, we have almost vertical drop 
of $\nu_{n_0,s_0}$ at $T^\infty\sim (2-4)\times 10^8\,\rm{K}$ (see Fig.\ \ref{Fig:inst})
%
\footnote{We do not plot the curves for an NS with $M=1.4M_\odot$ 
to avoid cluttering of the figure; they are very similar to those plotted for an NS with $M=1.8M_\odot$.}. 
%
The exception is the low-mass configurations in the model C, 
which have low values of $T_{{\rm c}n\, {\rm max}}^\infty$ (see Fig.\ \ref{Fig:Tc}). 
Dot-dashed line in the right panel
illustrates this point, 
corresponding
to the $n_0,s_0$ avoided-crossing for $M=1.1M_\odot$ NS in the model C.

The $n_0,s_1$ avoided-crossing 
occurs at
lower rotation frequencies than the $n_0,s_0$ one.
For the models A and B (representing wide critical temperature profiles) 
the corresponding $n_0,s_1$ curves 
pass 
through the sources in the instability window, 
and 
some stellar models (e.g., high-mass configurations for APR EOS),
allow one to explain not too hot sources. 
At the same time, in the model C, $n_0,s_1$ avoided-crossing lies at much lower rotation rates. 
This happens because in this model $T_{{\rm c}n}$ profile drops sharply in the outer core, 
shrinking the SF region even at low temperatures. 
As we checked for various SF models, such 
shrinking of SF region in the outer core
leads to a dramatic decrease of 
$\nu_{n_0,s_1}$ at a given $T^\infty$. 
Analogous shrinking of SF region due to drop of $T_{{\rm c}n}$ at its higher-density slope 
(in the stellar center) leads to the same effect, 
which is, however, not so dramatic.

\begin{figure}
    \begin{center}
        \leavevmode
        \includegraphics[width=7in]{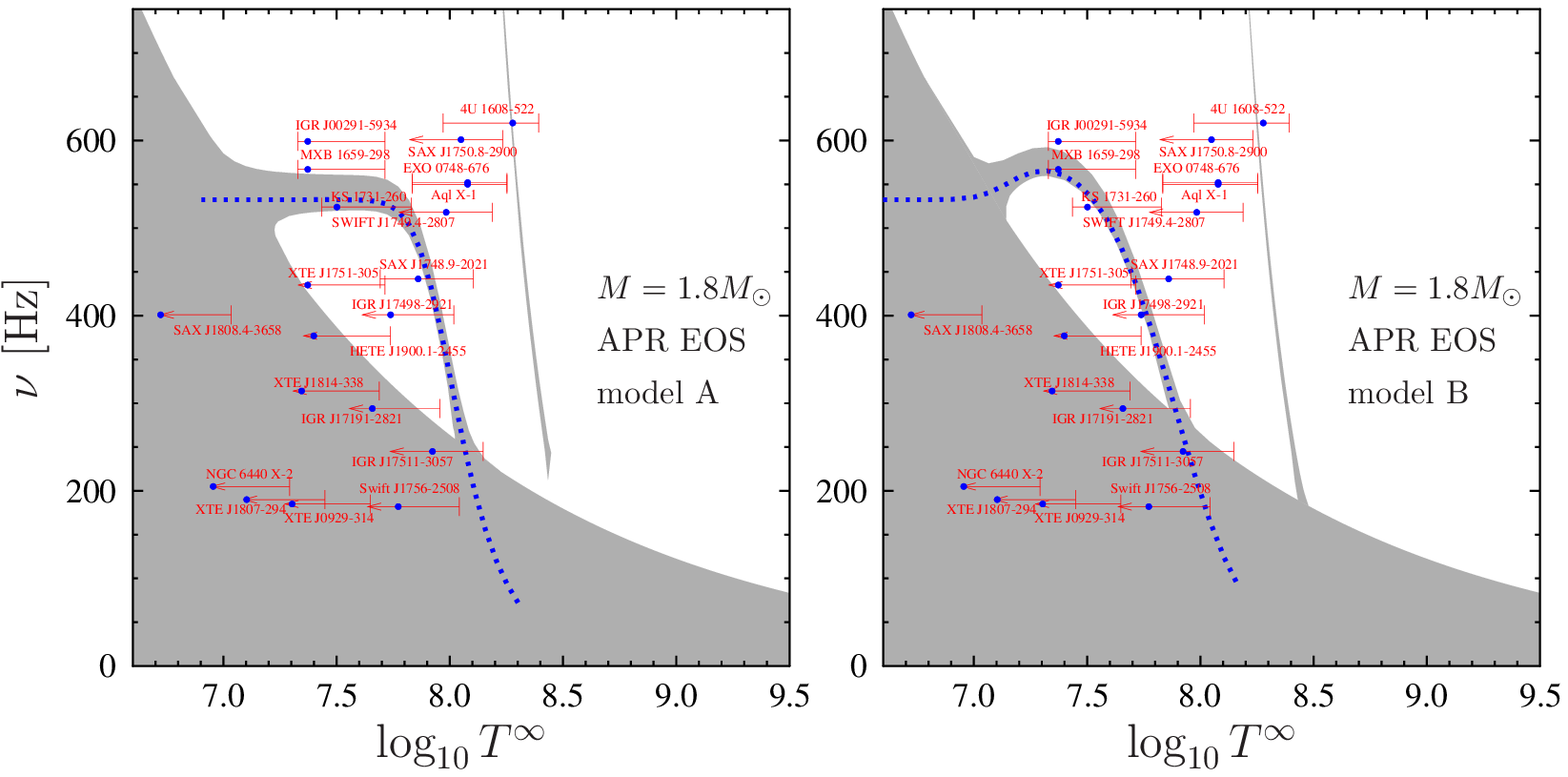}
    \end{center}
    \caption{Instability windows for $l=m=2$ normal nodeless r-mode calculated for $M=1.8M_\odot$ NS with 
    	APR EOS and the SF models A and B (left and right panels, respectively). 
    	In the region filled gray NS is stable.  
		At low frequencies high-temperature stability peak 
		in the model A is not calculated due to numerical problems. 
		Dotted lines show 
		$\nu_{n_0,s_1}(T^\infty)$ 
		(the same lines as in Fig.\ \ref{Fig:inst}). 
		We do not plot the curves $\nu_{n_0,s_0}(T^\infty)$ here,
		but they follow the corresponding stability peaks. See text and Fig.\ \ref{Fig:inst} for further details.
		}
    \label{Fig:peaks}
\end{figure}

In the vicinity of the avoided-crossings normal r-mode exhibits 
stabilizing resonance interaction with the SF r-modes \cite{gck14a,gck14b}. 
This leads to formation of ``stability strips'' 
along the $\nu_{n_0,s_\alpha}(T^\infty)$ curves in the $\nu-T^\infty$ plane 
(termed stability peaks in the initial scenario of Refs.\ \cite{gck14a,gck14b}). 
Fig.\ \ref{Fig:peaks} shows two examples of the instability windows 
with such strips for an NS with $M=1.8M_\odot$. 
The figure is plotted 
for the models A (left panel) and B (right panel),
assuming APR EOS.
To calculate these windows we accounted for the shear viscosity and mutual friction dissipation, 
as 
described in Ref.\ \cite{kg17}.
Note that our results imply that stability peaks are not vertical, 
as simple model of Refs.\ \cite{gck14a,gck14b} suggested. 
Nevertheless, according to the scenario of
Refs.\ \cite{gck14a,gck14b},
the star in the course of its evolution in LMXB 
should still spend most of its life climbing up the left edge of the ``peak''.
Thus, all the observed sources should be located 
within the stability region, which, however, varies with the stellar mass.

\section{Discussion and conclusions}
\label{sec:conc}

In order to confirm the viability of the phenomenological 
scenario
of r-mode resonance stabilization suggested in Refs.\ \cite{gck14a,gck14b} 
and to put it on a solid ground, 
we have calculated temperature-dependent r-mode spectrum for a slowly rotating SF NS. 
In contrast to the previous studies \cite{kg17,dkg19}, 
we, for the first time, 
accounted for both entrainment between neutrons and protons ($Y_{np}\neq 0$)
and the presence of muons in the NS core. 
Both muons and entrainment affect the spectrum qualitatively. 
Accounting for these effects together leads to elimination of rotational
SF modes 
from the oscillation spectrum, 
if one works {\it in the leading order} in rotation and assumes the standard ordering of eigenfunctions in $\Omega$.
This unphysical elimination takes place because non-vanishing entrainment 
requires non-vanishing radial displacement of the normal fluid component, 
while stratification by muons 
forbids such displacements
in the leading order in rotation. 
To restore SF r-modes and calculate the spectrum 
we developed an original perturbation scheme, 
whose leading order corresponds to the leading order in rotation and vanishing entrainment, while the next-to-the-leading order includes corrections due to non-vanishing entrainment (which is treated as a small parameter) and rotation {\it simultaneously}.

Using this perturbation scheme we proposed a simple method to calculate the eigenfrequency of SF r-modes [namely, the first correction to the known leading-order value $\sigma_0=2/(m+1)$] 
in the limit of vanishing rotation rate. 
We demonstrated that it can be calculated from the simple algebraic equation (\ref{condOmega0}).  

Then, applying this scheme to the
more general situation when the rotation frequency is small but non-vanishing, 
we have calculated temperature-dependent r-mode spectrum of an NS 
for two EOSs and three SF models (A, B, and C). 
We found that the normal nodeless $l=m=2$ r-mode 
(that is, the most unstable one \cite{ak01}) 
exhibits
avoided-crossings with the SF r-modes. 
Near the avoided-crossings normal r-mode experiences stabilizing 
interaction with SF modes, 
which suppresses the r-mode instability. 
When the avoided-crossing takes place in the region of the classical instability window, it results in the stability strip in the ``stellar temperature -- rotation frequency'' plane (termed ``stability peak'' in the initial phenomenological 
scenario
\cite{gck14a,gck14b}).
Although the strips, calculated here, are not vertical
as suggested in Refs.\ \cite{gck14a,gck14b},
an NS in LMXB will spend the majority of time climbing up the left 
edge of such strip \cite{gck14a,gck14b,kgc16,cgk17},
according to the resonance stabilization scenario.
Our calculations demonstrate that for certain SF models 
avoided-crossings take place in the range of parameters relevant to NSs in LMXBs, 
falling within the classical instability window (see Fig. \ref{Fig:inst}). 
This puts the scenario of Refs.\ \cite{gck14a,gck14b} on a solid ground,
making it a quantitative theory, 
which can explain observations
of hot, rapidly rotating NSs in LMXBs.

Our calculations open up a possibility to constrain parameters of neutron SF.
First, one can estimate $T_{{\rm c}n\, {\rm max}}$ by noticing that the hottest sources can be stabilized by the resonance with the main harmonic of SF r-mode ($n_0,s_0$).
For both considered EOSs, all the stellar masses, 
and all SF models considered by us
(not only the models A, B, C discussed in the paper) 
the curve $\nu_{n_0,s_0}(T^\infty)$
rapidly drops to zero at 
$T^\infty=T^\infty_{{\rm c}n\, {\rm max}}$.
In order for the curve to pass through the hottest sources in Fig.~\ref{Fig:inst},
one should require $T_{{\rm c}n\, {\rm max}}\sim (3-6)\times 10^8\,\rm K$ (see Ref.\ \cite{kgd20} for details).
Note that this estimate is the {\it lower limit} for the maximal $T_{{\rm c}n}$,
since $T_{{\rm c}n\, {\rm max}}$ denotes the maximum of $T_{{\rm c}n}$ 
in the region, 
where both 
neutrons and protons are SF.
Our constraint is consistent with microscopic calculations \cite{ls01, yls99,gps14,dlz14, drddwcp16, sc19},
as well as with observations of cooling NSs 
\cite{gkyg04,plps04,gkyg05,syhhp11,page11,CasA13,CasA15,bhsp18}.

Second, the resonance with the first overtone of SF r-modes ($n_0,s_1$)
can stabilize less hot rapidly rotating NSs
for our models A and B (wide neutron critical temperature profiles, 
most of the stellar core is SF), but not for the model C (narrow profile).
In the case of narrow $T_{{\rm c}n}$ profiles, resulting in much lower values of $\nu_{n_0,s_1}$,
these NSs can be stabilized only by the ($n_0,s_0$) resonance, if they have sufficiently low masses
(see dot-dashed line in the right panel of Fig.\ \ref{Fig:inst}).
However, this explanation is less plausible since NSs in LMXBs are generally believed to have high masses \cite{ozel12,antoniadis16}.
It is also worth mentioning that if
$T_{{\rm c}n\, {\rm max}}>10^9\,\rm K$ and $T_{{\rm c}n}$ profile is wide, 
i.e., $T_{{\rm c}n}$ remains high in the whole NS core, then hottest rapidly rotating sources, 
could 
be stabilized 
by the resonance with the first overtone of SF r-mode. 
Unfortunately, in this case we do not see a possibility to stabilize other, 
moderately heated, sources within our minimal scenario.

Resuming, to explain all wealth of observational data on NSs in LMXBs within the 
scenario of resonance r-mode stabilization 
\cite{gck14a,gck14b} (which we consider as a minimal  extension of the classical scenario), 
one needs a wide neutron critical temperature profile (see above), 
such that in the region of co-existence of neutron and proton SFs the maximum value of $T_{{\rm c}n}$ is $T_{{\rm c}n\, {\rm max}}\sim (3-6)\times 10^8\,\rm K$. 
The real maximum of $T_{{\rm c}n}$ throughout the whole density range 
can be larger than $T_{{\rm c}n\, {\rm max}}$. 
Note that proton SF model
cannot be constrained 
in our scenario
since 
proton pairing
only weakly affects the oscillation modes \cite{ga06,gkcg13,gkgc14}.

In principle, the positions of stability strips are sensitive not only to the neutron SF model, but also to EOS: as one can see from Fig.~\ref{Fig:inst}, APR and BSk24 EOSs
yield the results that differ by tens of percent.
However, in order to obtain constraints on EOS based on the proposed mechanism,
one has to calculate the r-mode spectrum more accurately, 
by
allowing for
gravitational field perturbations, higher-order terms in the expansions (\ref{expan1})--(\ref{expan2}), as well as
the General Relativity effects.

Definitely, one should also keep in mind that the real instability window 
may also be affected by additional r-mode stabilization mechanisms
such as Ekman layer dissipation
\cite{lu01,glama06},
bulk viscosity in hyperon/quark matter
\cite{no06,alford12,oghf19},
enhanced mutual friction dissipation \cite{hap09} etc., which are not considered in the present paper.
Accounting for all these effects can, in principle, modify our constraints on the parameters of neutron SF.

\section*{Acknowledgments}
This work is supported by the Russian Science Foundation (grant number 19-12-00133).

\appendix

\section{Constructing a solution for superfluid modes in the limit of vanishing rotation rate}

To start with, we again, 
for simplicity, consider a two-layer star 
consisting of the SF $npe\mu$ core and the crust.
More realistic structure with $npe$-layer in between 
is discussed in the end of this Appendix, 
and analyzed by us numerically as well. 
It shows the same qualitative behavior as the simplified two-layer stellar model. 

Let us
try to
build a solution for SF modes in the limit $\Omega \rightarrow 0$. 
Integrating oscillation equations 
in the crust, 
we find eigenfunctions 
$\xi_{{b} r, m+1,m}^{1}$ and $T^0_{b\;mm}$, 
which have a standard ordering in the crust: 
$\xi_{{b} r, m+1,m}^{1}\sim \Omega^2 T^0_{b\;mm}$. 
Generally, 
these functions do not vanish at the core-crust interface ($r=R_{\rm cc}$). 
This means that, due to the boundary conditions at the interface 
(continuity of $\xi_{{b} r, m+1,m}^{1}$ and $T^0_{b\;mm}$), 
$\xi_{{b} r, m+1,m}^{1}$ and $T^0_{b\;mm}$ 
should also be ordered as
$\xi_{{b} r, m+1,m}^{1}\sim \Omega^2 T^0_{b\;mm}$ 
in the core at $r=R_{\rm cc}$. 
However, as it was shown in Sec.\ \ref{zerolimit}, 
$\xi \sim \Omega T$. 
Generally this means that, in the SF $npe\mu$ core, 
$\xi_{{b} r, m+1,m}^{1}\sim \Omega T^0_{b\;mm}$ [see Eqs.\ (\ref{xidef}) and (\ref{Tdef})], and
as a result, the boundary conditions at the core-crust interface cannot be satisfied.

There is, however, a possibility to solve this problem.
The idea is to construct such a solution in the core that $\xi$ and thus $\xi_{{b} r, m+1,m}^{1}$ 
(note that $\xi=\xi_{{b} r, m+1,m}^{1}$ at $r=R_{\rm cc}$) 
is suppressed by a factor of $\Omega$ (or, in other words, almost vanish) at $r=R_{\rm cc}$,
while $T$ is not suppressed. 
Such solution satisfies the desired ordering, $\xi_{{b} r, m+1,m}^{1}\sim \Omega^2 T^0_{b\;mm}$. 
Continuity of $T^0_{b\;mm}$ can be ensured by an appropriate choice of the normalization constant,
while slightly varying $\sigma_1$ one can choose
$\xi_{{b} r, m+1,m}^{1}$ at $r=R_{\rm cc}$ 
in such a way
to satisfy the continuity of $\xi_{{b} r, m+1,m}^{1}$
at the core-crust interface.

To verify if we indeed can construct such a solution, 
we should solve Eq.\ (\ref{eq22}):
\begin{eqnarray}
	\frac{d^2 \xi}{dr^2}
	-\frac{K(r,\sigma_1)}{\Omega^2} \xi
	= 0.
	\label{eq22A}
\end{eqnarray}
Strictly speaking, Eq.\ (\ref{eq22A}) is not valid in the very vicinity of the stellar center 
(because when deriving it, we assumed that $r$ is not small). 
On the other hand, the asymptotic formulas, 
Eqs.\ (\ref{bccenter10})--(\ref{bccenter12}), 
are valid in the very vicinity of the center. 
Let us denote by $r_{\rm c}$  a coordinate 
such that the asymptotes (\ref{bccenter10})--(\ref{bccenter12}) 
are valid at $r\leq r_{\rm c}$, 
while Eq.\ (\ref{eq22A}) is valid at $r_{\rm c}\leq r\leq R_{\rm cc}$. 
The value of $r_{\rm c}$ can, in principle, 
be determined from Eqs.\ (\ref{Teq1})--(\ref{zeq1}) by analyzing behavior 
of various terms at $\Omega\rightarrow 0$ and $r\rightarrow 0$. 
In the limit of vanishing rotation rate $r_{\rm c}\rightarrow 0$.
In what follows, we make use of  the asymptotic formulas, 
Eqs.\ (\ref{bccenter10})--(\ref{bccenter12}) 
at $0\leq r\leq r_{\rm c}$,
and solve Eq.\ (\ref{eq22A}) at $r_{\rm c}\leq r\leq R_{\rm cc}$.

Note that Eq.\ (\ref{eq22A}) has the same form as the Schr\"{o}dinger equation. 
The limit $\Omega \rightarrow 0$ corresponds to the quasi-classical limit, in which the solution to Eq.\ (\ref{eq22A}) far from the ``turning points'' [where $K(r,\sigma_1)$ vanishes] can be written as \cite{ll65}
\begin{eqnarray}
	 \xi=\frac{A_1}{K(r,\sigma_1)^{1/4}}\;{\rm exp}\left[\frac{\int_{r_{\rm c}}^r \sqrt{K(\tilde{r},\sigma_1)}d\tilde{r}}{\Omega}\right]+
	\frac{A_2}{K(r,\sigma_1)^{1/4}}\;{\rm exp}\left[-\frac{\int_{r_{\rm c}}^r \sqrt{K(\tilde{r},\sigma_1)}d\tilde{r}}{\Omega}\right], \label{solcaseI}
\end{eqnarray}
where $A_1$ and $A_2$ are some constants.

Consider first a situation when $K(r,\sigma_1)>0$ in the whole SF $npe\mu$ core (at $r_{\rm c}\leq r\leq R_{\rm cc}$). Then we have no 
``turning points'' and the solution (\ref{solcaseI}) is valid throughout the core. 
Let us check if we can meet all the boundary conditions in this case.

The asymptotic solution (\ref{bccenter10})--(\ref{bccenter12})
implies that [see also Eqs.\ (\ref{xidef}) and (\ref{Tdef})]
\begin{eqnarray}
T=\frac{3+2m}{r_{\rm c}}\xi
\end{eqnarray}
at $0\leq r\leq r_{\rm c}$. 
Keeping in mind that $T=d \xi/dr$ [see Eq.\ (\ref{eq111})] at $r\geq r_{\rm c}$, 
and making use of the continuity of eigenfunctions $\xi_{{b} r, m+1,m}^{1}$, $z_{r, m+1,m}^{1}$, $T^0_{b\;mm}$, and $T^0_{z\,mm}$ (and thus the continuity of $T$ and $\xi$) at $r=r_{\rm c}$, we find
\begin{eqnarray}
\d{\xi}{r} =\frac{3+2m}{r_{\rm c}}\xi \label{relA}
\end{eqnarray}
at $r=r_{\rm c}$,
or in view of Eq.\ (\ref{solcaseI}):
\begin{eqnarray}
\frac{3+2m}{r_{\rm c}}(A_1+A_2)=(A_1-A_2)\frac{\sqrt{K(r_{\rm c},\sigma_1)}}{\Omega}-\frac{1}{4}(A_1+A_2)\frac{K'(r_{\rm c},\sigma_1)}{K(r_{\rm c},\sigma_1)}\approx (A_1-A_2)\frac{\sqrt{K(r_{\rm c},\sigma_1)}}{\Omega},\label{bc2}
\end{eqnarray}
where the prime denotes the derivative with respect to $r$.
One can see that the constants $A_1$ and $A_2$ are of the same order 
of magnitude.
Thus, at the core-crust interface the decreasing exponent 
in the solution (\ref{solcaseI}) will be negligibly small and we find 
\begin{eqnarray}
	 \xi=\frac{A_1}{K(R_{\rm cc},\sigma_1)^{1/4}}\;{\rm exp}\left[\frac{\int_{r_c}^{R_{\rm cc}} \sqrt{K(\tilde{r},\sigma_1)}d\tilde{r}}{\Omega}\right],\\
	T=\frac{A_1 K(R_{\rm cc},\sigma_1)^{1/4}}{\Omega}\;{\rm exp}\left[\frac{\int_{r_c}^{R_{\rm cc}} \sqrt{K(\tilde{r},\sigma_1)}d\tilde{r}}{\Omega}\right]
\end{eqnarray}
at $r=R_{\rm cc}$ with $\xi \sim \Omega T$ and no possibility to 
match
eigenfunctions 
at the core-crust interface (see above).

Assume now that $K(r,\sigma_1)$ vanishes at some radius, 
then the solution (\ref{solcaseI}) is invalid in the vicinity of this radius \cite{ll65} and consideration above is not applicable.
Consider a situation when $K(r,\sigma_1)$ has a negative minimum at some radius $r_0$ such that $r_{\rm c}<r_0<R_{\rm cc}$, $K(r_0,\sigma_1)=-K_0$.
The function $K(r,\sigma_1)$ can be expanded near $r_0$ as
\begin{eqnarray}
K(r,\sigma_1)=-K_0+\beta(r-r_{\rm 0})^2, \label{parabola}
\end{eqnarray}
with $\beta\equiv K''(r,\sigma_1)|_{r_0}/2>0$, and Eq.\ (\ref{eq22A}) takes the form
\begin{eqnarray}
	\frac{d^2 \xi}{dx^2}
	-\left(a+\frac{1}{4}x^2\right) \xi= 0,
	\label{eq22B}
\end{eqnarray}
where $a \equiv -K_0/(2\Omega\beta^{1/2})$, $x \equiv (4\beta)^{1/4}(r-r_0)/\sqrt{\Omega}$.
Note that, in the limit $\Omega \rightarrow 0$, $x$ changes from $-\infty$ to $+\infty$. 
The solution to Eq.\ (\ref{eq22B}) can be expressed in terms of parabolic cylinder functions \cite{as72}.
General solution at $x\geq 0$ can be presented as a linear combination of functions $U(a,x)$ and $V(a,x)$ introduced in Ref.\ \cite{as72},
$\xi=B_1 U(a,x)+B_2 V(a,x)$, and
exhibiting the following asymptotic behavior at $x\gg |a|$
\begin{eqnarray}
U(a,x)\approx{\rm exp}\left(-\frac{x^2}{4}\right)x^{-a-\frac{1}{2}},\label{asy1}\\
V(a,x)\approx \sqrt{\frac{2}{\pi}}{\rm exp}\left(\frac{x^2}{4}\right)x^{a-\frac{1}{2}}.
\end{eqnarray}
Notice that, since Eq.\ (\ref{eq22B}) is symmetric with respect to the transformation $x\rightarrow -x$, 
the functions $U(a,-x)$ and $V(a,-x)$ are also solutions to Eq.\ (\ref{eq22B}).
Since we are interested in real $\xi$,
we shall present the solution to Eq.\ (\ref{eq22B}) 
 in the region $x\leq 0$ 
as a linear combination of functions $U(a,-x)$ and $V(a,-x)$, $\xi=B_3 U(a,-x)+B_4 V(a,-x)$, 
with the following asymptotic behavior at $-x\gg |a|$
\begin{eqnarray}
U(a,-x)\approx{\rm exp}\left(-\frac{x^2}{4}\right)(-x)^{-a-\frac{1}{2}},\\
V(a,-x)\approx \sqrt{\frac{2}{\pi}}{\rm exp}\left(\frac{x^2}{4}\right)(-x)^{a-\frac{1}{2}}.\label{asy4}
\end{eqnarray}
The constants $B_1,B_2,B_3$, and $B_4$ should be found 
by matching the functions $\xi$ and $T$ in different regions described below.

Let us consider three regions. 
In the region I, $r_{\rm c}\leq r\leq r_{\rm I}$, 
$K(r,\sigma_1)$ is positive and $r_{\rm I}$ is sufficiently far from the turning point, 
so that quasi-classical approximation and hence the solution (\ref{solcaseI}) are valid. 
In the region III, $r_{\rm III}\leq r\leq R_{\rm cc}$, 
$K(r,\sigma_1)$ is also positive and the point $r_{\rm III}$ 
is again chosen far enough away from the turning point, 
so that 
the solution in this region 
reads
\begin{eqnarray}
	 \xi=\frac{C_1}{K(r,\sigma_1)^{1/4}}\;{\rm exp}\left[\frac{\int^{R_{\rm cc}}_r \sqrt{K(\tilde{r},\sigma_1)}d\tilde{r}}{\Omega}\right]+
	\frac{C_2}{K(r,\sigma_1)^{1/4}}\;{\rm exp}\left[-\frac{\int^{R_{\rm cc}}_r \sqrt{K(\tilde{r},\sigma_1)}d\tilde{r}}{\Omega}\right]. \label{solIII}
\end{eqnarray}
In region II, $r_{\rm I}\leq r\leq r_{\rm III}$, 
we assume that the expansion (\ref{parabola}) is valid, 
and, moreover, $|x|\gg a$
at $r=r_{\rm I}$ and $r=r_{\rm III}$.

Boundary condition at $r=r_{\rm c}$ prescribes the constants $A_1$ and $A_2$ to be of the same order 
of magnitude
(see Eq.\ \ref{bc2}). 
As a result, the eigenfunctions in the region I at $r=r_{\rm I}$ read 
(the decreasing exponents are neglected)
\begin{eqnarray}
	 \xi=\frac{A_1}{K(r_{\rm I},\sigma_1)^{1/4}}\;{\rm exp}\left[\frac{\int_{r_{\rm c}}^{r_{\rm I}} \sqrt{K(\tilde{r},\sigma_1)}d\tilde{r}}{\Omega}\right], 
	 \label{xi}\\
	T=\frac{d\xi}{dr}=A_1\frac{ K(r_{\rm I},\sigma_1)^{1/4}}{\Omega}\;{\rm exp}\left[\frac{\int_{r_{\rm c}}^{r_{\rm I}} \sqrt{K(\tilde{r},\sigma_1)}d\tilde{r}}{\Omega}\right]. \label{TApp}
\end{eqnarray}
Similarly, the boundary condition at the core-crust interface, $\xi(R_{\rm cc})=0$ (see above), 
requires $C_1=-C_2$. 
Thus, the eigenfunctions in the region III at $r=r_{\rm III}$ equal
(we again omit the small
exponent $\propto C_2$)
\begin{eqnarray}
	 \xi=\frac{C_1}{K(r_{\rm III},\sigma_1)^{1/4}}\;{\rm exp}\left[\frac{\int_{r_{\rm III}}^{R_{\rm cc}} \sqrt{K(\tilde{r},\sigma_1)}d\tilde{r}}{\Omega}\right], 
	 \label{xi2}\\
	T=\frac{d\xi}{dr}=C_1\frac{ K(r_{\rm III},\sigma_1)^{1/4}}{\Omega}\;{\rm exp}\left[\frac{\int_{r_{\rm III}}^{R_{\rm cc}} \sqrt{K(\tilde{r},\sigma_1)}d\tilde{r}}{\Omega}\right]. \label{T2}
\end{eqnarray}
Further, using the asymptotic behavior (\ref{asy1})--(\ref{asy4}) of the eigenfunctions in the region II and continuity of eigenfunctions $\xi$ and $T$ at $r=r_{\rm I}$ and $r=r_{\rm III}$ we find that $B_2\ll B_1$ and $B_4\ll B_3$.

At $x=0$ eigenfunctions also have to be continuous, which means
\begin{eqnarray}
B_1 U(a,x)=B_3 U(a,-x),\\
B_1 \frac{dU(a,x)}{dx}=-B_3 \frac{dU(a,-x)}{dx}
\end{eqnarray}
at $x\rightarrow 0$.
The above system has two solutions. 
First, $U(a,0)=0$, $B_1=-B_3$. 
Second, $d U(a,x)/dx|_{x=0}=0$, $B_1=B_3$. 
Since (e.g., \cite{as72})
\begin{eqnarray}
U(a,0)=\frac{\sqrt{\pi}}{2^{\frac{1}{2}a+\frac{1}{4}}\Gamma \left(\frac{3}{4}+\frac{1}{2}a\right)},\\
\left.\frac{d U(a,x)}{dx}\right|_{x=0}=-\frac{\sqrt{\pi}}{2^{\frac{1}{2}a-\frac{1}{4}}\Gamma \left(\frac{1}{4}+\frac{1}{2}a\right)},
\end{eqnarray}
the first and the second solutions correspond to, respectively, 
$\frac{3}{4}+\frac{1}{2}a=-n$ and $\frac{1}{4}+\frac{1}{2}a=-n$, 
where $n$ is a natural number. 
Merging these constraints, we find the ``quantization rule'' for the parameter $a$: 
$a=-\frac{1}{2}-n$. 
Recalling the definition of $a$, we get
\begin{eqnarray}
K_0=2\Omega\sqrt{\beta}\left(\frac{1}{2}+n\right). \label{constr}
\end{eqnarray}
This indirect condition on the frequency $\sigma_1$ 
can also be rewritten in the form of the Bohr-Sommerfeld quantization rule as
\begin{eqnarray}
\int_{r_1}^{r_2}\sqrt{-K(r,\sigma_1)} \, dr=\pi\Omega\left(\frac{1}{2}+n\right),
\end{eqnarray}
where $r_1$ and $r_2$ are the turning points, $K(r_1,\sigma_1)=K(r_2,\sigma_1)=0$.
It is interesting to note that  the condition 
(\ref{constr}) 
could be immediately obtained from the analogy to 
the quantum harmonic oscillator problem \cite{ll65}. 
This is not surprising, 
since Eq.\ (\ref{eq22B}) describes harmonic oscillations, 
while the boundary conditions imposed in the region II  
require vanishing of the ``wave-function'' $\xi$ at $x\rightarrow \pm \infty$.
Therefore, this problem is indeed completely equivalent to the 
classical problem of quantum mechanics,
and leads to the same spectrum (\ref{constr}).
As in quantum mechanics, 
$n$ in Eq.\ (\ref{constr}) determines the number of zeros of the function $\xi$
and its parity.

Now the asymptotic solution in the region II at $|x|\gg |a|$ 
can be represented as
\begin{align}
&
\xi=B_1 \, U(a,x), \quad\quad x\gg |a|,
&
\nonumber\\
&
\xi=(-1)^n \, B_1\,  U(a,-x), \quad\quad -x\gg |a|.
&
\label{II}
\end{align}
This asymptotic solution should be matched with the solution (\ref{xi})--(\ref{TApp}) 
in the region I at $r=r_{\rm I}$ (where $-x\gg |a|$)
and with the solution (\ref{xi2})--(\ref{T2}) in the region III at $r=r_{\rm III}$ (where $x\gg |a|$).
In this way one can relate the coefficients $A_1$, $B_1$, and $C_1$, 
which completes our solution.
We leave this exercise for the reader.

Above we demonstrated how to build 
a solution for superfluid r-modes in the limit $\Omega \rightarrow 0$.
Our results imply 
that 
one needs to have a region with $K(r,\sigma_1)<0$ in the SF $npe\mu$ core. 
This region has to be small, $r_2-r_1\sim \sqrt{\Omega}$, 
in order for the solution to possess a finite number of nodes, $n$. 
In other words, in the limit $\Omega \rightarrow 0$, $\sigma_1$ for SF r-modes is set by the condition
\begin{equation}
{\rm min}[K(r,\sigma_1)]=0, \label{condA}
\end{equation} 
i.e., the minimal value of $K(r,\sigma_1)$ throughout the SF $npe\mu$ core must vanish. 
Obviously, $\sigma_1$ for various harmonics (with finite number of nodes) must 
coincide in this limit. 
The consideration above demonstrates that r-mode eigenfunctions cannot be expanded in Taylor series in $\Omega$ at $\Omega \rightarrow 0$ (see Eqs.\ \ref{solcaseI} and \ref{solIII}), 
because they are {\it non-analytic} 
in this limit.

We assumed above that minimum value of $K(r,\sigma_1)$ 
corresponds to a minimum of the function $K(r,\sigma_1)$, that is $K'(r_0,\sigma_1)=0$. 
However, $K(r,\sigma_1)$ can reach its minimum value at the stellar center or at the core-crust interface, where $K'(r,\sigma_1)$, generally, does not vanish. 
Considering this situation in a way similar to that described above, 
one can easily check that our main conclusions, in particular, Eq.\ (\ref{condA}), remain valid also in this case. 

Let us now turn to the more realistic case of a three-layer star,
containing $npe$ layer in the outer core,
between the inner $npe\mu$ core and crust. 
We limit ourselves to considering two possibilities. 
First, neutrons are SF all the way from the stellar center 
to some point $r=r_{\rm s}$ {\it inside} the $npe\mu$ core. 
Then at $r>r_{\rm s}$ neutrons are normal and eigenfunctions obey the standard ordering at $r_{\rm s}\leq r\leq R$, 
$\xi_{{b} r, m+1,m}^{1}\sim \Omega^2 T^0_{b\;mm}$. 
Hence all the reasoning presented above remains unchanged in this case, with the only difference, that now one should vanish $\xi$ at $r=r_{\rm s}$: $\xi(r_{\rm s})=0+O(\Omega)$.

The second possibility is that neutron SF extends from the stellar center 
up to some point $r=r_{\rm s}$ {\it outside} the $npe\mu$ core, $r_{\rm s}>R_\mu$.
This situation differs slightly from that analyzed above. 
Now, while in the crust and in the normal part of the $npe$ core $\xi_{{b} r, m+1,m}^{1}\sim \Omega^2 T^0_{b\;mm}$, in the SF $npe$ matter $\xi_{{b} r, m+1,m}^{1}\sim z_{r, m+1,m}^{1}\sim T^0_{b\;mm}\sim T^0_{z\,mm}$ \cite{dkg19}. 
Thus, in this situation (and in the limit $\Omega \rightarrow 0$) not $\xi$, but the function $T$ should vanish in the SF $npe\mu$ core interface, at $r=R_\mu$, to guarantee the continuity of eigenfunctions at $r=R_\mu$. 
Nevertheless, even in this case the conclusions of this Appendix 
[in particular, Eq.\ (\ref{condA})] remain unaffected.

\end{document}